\documentclass[journal]{IEEEtran}
\usepackage{cite}

\ifCLASSINFOpdf
   \usepackage[pdftex]{graphicx,epsfig}
\else
   \usepackage[dvips]{graphicx}
\fi
\usepackage[cmex10]{amsmath}
\usepackage[subnum]{cases}

\usepackage{algorithm}
\usepackage{algorithmic}

 \usepackage[dvips]{graphicx}
\usepackage[tight,footnotesize]{subfigure}

\ifCLASSOPTIONcompsoc
\usepackage[tight,normalsize,sf,SF]{subfigure}
\else
\usepackage[tight,footnotesize]{subfigure}
\fi

\begin{document}
%
\title{An analysis of transient impulsive noise in a Poisson field of interferers for wireless channel in substation environments}
%
%
%

\author{Minh~Au,~\IEEEmembership{Student Member,~IEEE,}
        Basile~L.~Agba,~\IEEEmembership{Senior~Member,~IEEE,}
        and~Fran\c cois~Gagnon,~\IEEEmembership{Senior~Member,~IEEE}
\thanks{M. Au is PhD student at Lacime laboratory in Electrical Engineering Department
at \'Ecole de technologie sup\'erieure, Montr\'eal,
QC, H3C1K3 e-mail: Minh.au@lacime.etsmtl.ca.}
\thanks{B.L. Agba is PhD and researcher at IREQ (Institut de recherche Hydro-Qu\'ebec and F. Gagnon is professor in Electrical Engineering Department at
\'Ecole de technologie sup\'erieure.}
}

\maketitle

\begin{abstract}
In substations, the presence of random transient impulsive interference sources makes noise highly non-Gaussian. In this paper, the primary interest is to provide a general model for wireless channel in presence of these transient impulsive noise for space-time signal processing problems. We assume a superposition of independent interference sources randomly distributed in space-time in a Poisson field of interferers. By using stochastic geometry approach, first order and second order statistics can be derived from basic waveforms of impulsive interferers. We use discrete-time series model to simulate the random transient impulsive waveforms. It is demonstrated that the amplitude distribution and density of the proposed model converges to $\alpha$-stable distributions and their power spectral densities are $\sim 1/(f-f_{0})^{k}$ where $f_{0}\geq 0$ is a resonant frequency and $k > 0$. Measurements and computer simulations are provided where impulsive noise are to demonstrate the efficiency of the analysis. 
\end{abstract}


%
\IEEEpeerreviewmaketitle

\section{Introduction}
\label{sect:I}
\IEEEPARstart{T}{he} presence of interferences in most environments invalid the Gaussian noise model. For example, in the presence of impulsive interferences, the noise model is non-Gaussian \cite{Wegman1989,Middleton1999,Middleton1977}. Their impact on communication systems performances can be severely degraded \cite{Middleton1999,Spaulding1977,Blum1999}. This paper is particularly focused on impulsive noise in substation environments. They include electrical breakdown discharges phenomena such as partial discharges (PD), electrical arcs in addition to background noise. The induced radiations are transient impulsive waveforms. They can occupy a wide frequency range which interfere with conventional wireless communication systems \cite{Madi2010,Ndo2013,Bhatti2009}. 

Impulsive noise modelling is an active research for the design of robust receivers in these environments. One of the most commonly used is Middleton class A model \cite{Spaulding1977,Madi2010,Bhatti2009,Middleton1973,Middleton1977}. It is a Poisson-Gaussian noise process where independent emissions of these impulses are assumed. In \cite{Middleton1999,Middleton1977} Middleton classifies transient impulsive noise in the Class B model. However, it may be challenging to simulate the noise model due to the complex form of the amplitude distribution and density which considers six parameters. An alternative approach can be proposed by using the well known $\alpha$-stable distributions which is an approximation of the Middleton Class B \cite{Middleton1999,Tsihrintzis1996}. In \cite{Chambers1976,Weron1995}, authors provide simulation methods to produce stable random variables. These models give suitable approximation of first order statistics due to strong amplitude and rare events of impulsive noises. In practice, the simulation results cannot produce transient effects induced by electrical discharges in substations. Thus, they are limited in terms of second order statistics. On the other hand, Markov-chains has been investigated for impulsive noise modelling in \cite{Zimmermann2002,Ndo2013}. States of the Markov-chain take into account various physical phenomena such as inter-arrival time, duration of impulses and their amplitude distributions. Nevertheless, the definition of number of states can be challenging and first and second order statistics cannot be written explicitly. 

In space-time signal processing problems, it is useful to consider the space-time distribution of interference sources for interference mitigation techniques and for communication theory in non-Gaussian noise \cite{Vaseghi2008,Middleton2012,Blum1999}. In this paper, we develop a new approach to achieve a non-Gaussian noise model in presence of impulsive interference sources where first order and second order statistics can be derived and written explicitly. We develop an original approach for random transient impulsive noise waveform modelling based on discrete-time series. Then, by using stochastic geometry approach, the first order of characteristic function of the random Poisson field of interferers can be derived. It is widely used in wireless communication for co-channel interference modelling \cite{Yang2003a,Gulati2010}, random access systems \cite{Llow1998} where statistics can be derived. Inspired by prior works published in \cite{Rice1944,Lowen1990,Lowen1989,Middleton1974,Yang2003a,Gulati2010,Llow1998,Carson1931}, we derive first and second order statistics respectively from Campbell's and Carson's theorems in tractable forms.

The paper is organized as follows. In section \ref{sect:III}, we provide a mathematical formulation of multiple interference sources in substation environments. The general waveform of impulsive noise is discussed and how the non-Gaussian noise process can be formulated in terms of first and second order statistics. In section \ref{sect:IV}, basic impulsive waveforms model is specified to derive the first order of characteristic function. Thus, we define a general impulsive waveform based on discrete-time series model. In section \ref{sect:V}, the non-Gaussian process is described in terms of first order statistics \textit{e.g}. moments-cumulants, amplitude distributions and densities. It is also described in terms of second order statistics \textit{e.g.} power spectral densities. Finally, in section \ref{sect:VI}, computer simulations and measurements are provided to demonstrate the efficiency of the analysis over computer simulations and measurements results.



\section{Mathematical formulation of multiple interference sources in substations}
\label{sect:III}

\subsection{Basic Poisson field of interferers}
\label{ssect:V.C}
Stochastic geometry approach is used to derive the first order of the characteristic function of interfering sources in presence of impulsive noise. Mathematical notations are inspired by \cite{Baccelli2009}. First, we need to define the statistics of the interference field. Let $\Pi(\mathbf{r},t)$ be a random space-time field as a linear superposition of individual fields randomly produced by activated sources (emission of radiations) in a domain of sources. We postulate the Poisson random field $ \Pi(\mathbf{r},t) = \sum_{i} \delta_{\mathbf{r}_{i},t_{i}} $ where $\delta_{\mathbf{r},t}$ is the Dirac measure on a finite space-time domain $\Lambda \subset \mathcal{R}^{3}\times \mathcal{R}$ where the three dimensional space is considered. We note that $\boldsymbol\Psi = \left\lbrace \mathbf{r},t \right\rbrace_{i} $ is a set of points representing active sources in the space-time domain $\Lambda$. Thus, the interference field can be written as :

\begin{equation}
\Pi(\mathbf{r},t) =  \sum\limits_{\psi \in \boldsymbol\Psi}  \boldsymbol\Upsilon_{\mathbf{r},t\vert  d(\mathbf{r},\mathbf{r}_{0})}
\label{eq:V.C.1}
\end{equation}
where $\boldsymbol\Upsilon \in \mathcal{L}^{2}(\Omega,\mathcal{F},P)$ is an ensemble of measurable random waveforms of emitting sources, $d(\mathbf{r},\mathbf{r}_{0})$ is a distance parameter where $\mathbf{r}_{0}$ is the position of the antenna and $t\vert d(\mathbf{r},\mathbf{r}_{0})$ is related to the delay of propagation induced by the position of an interference source and the receiver. We now assume that these impulsive noise are separable functions such that the interference field can be written as :

\begin{equation}
\Pi(\mathbf{r},t) = \sum\limits_{\mathbf{r} \in N_{\mathbf{r}}}  \sum\limits_{t \in N_{t}}\boldsymbol\Upsilon_{\mathbf{r}}\boldsymbol\Upsilon_{t\vert  d(\mathbf{r},\mathbf{r}_{0})}
\label{eq:V.C.2}
\end{equation}
where $N_{\mathbf{r}}$ is the point process related to emitting sources and $N_{t}$ is the point process of radiations in time domain. The intensity of measure of the point process $\mathcal{Z}\left( B\right) = E\left[  \boldsymbol\Psi(B)\right]$, where $B$ is a Borel set, has a density $\lambda(\psi) = \lambda(\mathbf{r},t)$. By using Campbell's theorem, the mean  interference field is :

\begin{equation}
E\left[ \Pi(\mathbf{r},t) \right] = \int_{\Lambda}  \boldsymbol\Upsilon_{\mathbf{r},t\vert  d(\mathbf{r},\mathbf{r}_{0})} \lambda(\psi)d\psi
 \label{eq:V.C.3}
 \end{equation} 
By assuming the ergodicity of the ensemble $\boldsymbol\Upsilon$ such that :

\begin{equation}
\left\langle \boldsymbol\Upsilon_{\mathbf{r},t\vert  d(\mathbf{r},\mathbf{r}_{0})}\right\rangle  = U_{\mathbf{r}}U_{t\vert  d(\mathbf{r},\mathbf{r}_{0})}
 \label{eq:V.C.5.bis}
\end{equation}
where $U_{\mathbf{r},t}$ is the basic waveform of interference sources. From the Laplace functional of the equation (\ref{eq:V.C.3}), the first order of characteristic function $Q(j\xi)$ of the superposition of these emitting radiations is given by :

\begin{equation}
Q(j\xi) = \exp\left(- \int_{\Lambda}\left\lbrace 1- \exp \left[ -j\xi U_{\mathbf{r},t} \right] \right\rbrace \lambda(\psi) d\psi\right) 
 \label{eq:V.C.4}
 \end{equation} 
 
The first order statistics such as moments, cumulants, amplitude distribution and density can be derived from the characteristic function which depends on the definition of the basic waveform of interference sources. It will be specified for impulsive interference sources in substation environments. The first order statistics of the interfering sources can be extended by considering an additive background noise. 

\subsection{Interference sources in substation}
\begin{figure}[!htbp]
\centering
\includegraphics[width=3.47in]{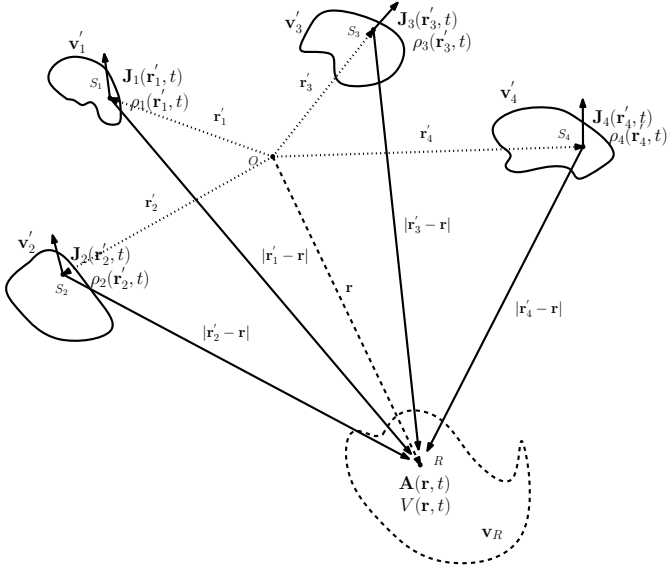}
\caption{Geometry of interfering sources and the receiver}
\label{fig:III.0.1}
\end{figure}

In substation environments, radiations from interferers received at the antenna are impulsive and are caused by partial discharges mainly in air. They can be located in HV equipments when physical conditions are reached to discharge such as presence of defects, high electric field, free electrons etc, \cite{Bartnikas1993}. In presence of multiple interference sources in the vicinity of the antenna, a low density of space Poisson process $N_{\mathbf{r}}$ can be assumed where impulsive sources are randomly located in the three dimensional space in far-field region. For the activated impulsive sources, charge particles and currents produced by a discharge radiate impulsive electromagnetic radiations. The fields $\mathbf{E}$ and  $\mathbf{B}$ can be obtained from retarded potentials $\mathbf{A}$ and $V$ respectively the magnetic potential vector and the scalar potential by :

\begin{subequations}
\begin{align}
\mathbf{E} & = - \nabla V - \frac{\partial \mathbf{A}}{\partial t}\\
\mathbf{B} & = \nabla \times \mathbf{A}
\end{align}
\label{eq:III.0.1}
\end{subequations}
The retarded vector and scalar potentials satisfying the Lorenz gauge condition can be written as a wave equation :

\begin{subequations}
\begin{align}
\nabla^{2} V - \frac{1}{c^{2}} \frac{\partial V}{\partial t} & = -\frac{1}{\varepsilon_{0}}\sum\limits_{\mathbf{r}\in N_{\mathbf{r}}}\rho(\mathbf{r},t)\\
\nabla^{2} \mathbf{A} - \frac{1}{c^{2}} \frac{\partial \mathbf{A}}{\partial t} & =  -\mu_{0}\sum\limits_{\mathbf{r}\in N_{\mathbf{r}}}\mathbf{J}(\mathbf{r},t)
\end{align}
\label{eq:III.0.2}
\end{subequations}
where $\rho(\mathbf{r},t)$ and $\mathbf{J}(\mathbf{r},t)$ are charge density and current density respectively of an activated sources $S_{i}$ on the emitting element $d\mathbf{v}^{'}_{i}$ in $\mathbf{v}^{'}_{i}$ as depicted on Fig. \ref{fig:III.0.1}. The sum $\sum_{\mathbf{r}\in N_{\mathbf{r}}}$ represents the superposition of individual source randomly located in the vicinity of the antenna $R$. The speed of light in the medium is represented by $c$, $\varepsilon_{0}$ and $\mu_{0}$ are the permittivity and the permeability of the vacuum respectively. Each activated source $S_{i}$ radiates electromagnetic waves in the medium induced by retarded potentials. A receiver $R$ can received these waves on the receiving element $d\mathbf{v}_{R}$ in $\mathbf{v}_{R}$. By considering successive radiations from each source in the time domain, the Poisson process should be extended to space-time process. 

Solution of equations in (\ref{eq:III.0.2}) is given by :

\begin{subequations}
\begin{align}
V(\mathbf{r},t) & = \frac{1}{4 \pi \varepsilon_{0}}\sum\limits_{\mathbf{r}^{'}\in N_{\mathbf{r}}}\int_{\mathbf{v}^{'}}\frac{\rho(\mathbf{r}^{'},t-\vert \mathbf{r}^{'}-\mathbf{r}\vert/c)}{\vert \mathbf{r}^{'}-\mathbf{r}\vert}d\mathbf{v}^{'}\\
\mathbf{A}(\mathbf{r},t) & = \frac{\mu_{0}}{4 \pi}\sum\limits_{\mathbf{r}^{'}\in N_{\mathbf{r}}}\int_{\mathbf{v}^{'}}\frac{\mathbf{J}(\mathbf{r}^{'},t-\vert \mathbf{r}^{'}-\mathbf{r}\vert/c)}{\vert \mathbf{r}^{'}-\mathbf{r}\vert}d\mathbf{v}^{'}
\end{align}
\label{eq:III.0.3}
\end{subequations}

The interference sources are independent such that radiations at the antenna surface is a superposition of independent impulsive noise. Generally, the receiver has a directional radiation pattern not necessarily omnidirectional. The antenna has an effective length related to the induced voltage or current at the terminals to the incident field $\mathbf{E}$. The receiver may also have RF and IF (intermediate frequency) stages \textit{e.g}, low noise amplifier and linear filters. As a result, the superposition of these radiations gives the typical waveform obtained from the receiver $R$ by :

\begin{equation}
\begin{split}
I_{t} &= \int_{\mathbf{v}_{R}} a_{R}(\theta,\zeta,t) \ast \mathbf{E}(\theta,\zeta,t) d\mathbf{v}_{R}\\
 &= \sum\limits_{t\in N_{t}} U_{t}\\
\end{split}
\label{eq:III.0.4}
 \end{equation} 
where $a_{R}(\theta,\zeta,t)$ is the aperture weighting function in spherical coordinate system $(r,\theta,\zeta)$ respectively represented by the radial distance, the polar and the azimuthal angles. The aperture weighting function includes both the radiation pattern of the antenna and the linear impulse response of filters. It can be seen as an impulse response of the receiver where the receiving field is converted into a time waveform alone. The convolution product operates for temporal impulse response. The antenna receives the electric field $\mathbf{E}$, induced by the activated interferers, on the receiving element $d\mathbf{v}_{R}$ in $\mathbf{v}_{R}$. The resulting waveform $I_{t}$ is a superposition of independent impulsive noise $U_{t}$ produced by activated interference sources. The process $I_{t}$ is excited by a Poisson process $N_{t}$ related to the number of impulses in time domain. It is denoted as a shot-noise process \cite{Rice1944,Gilbert1960}. The typical impulsive noise $U_{t}$ after any RF and IF stages of (linear) filtering is written as :

\begin{equation}
\begin{split}
U_{t} &= \frac{1}{4\pi}\bigg\Vert\frac{u(\theta,\zeta)}{r}\bigg\Vert u(t)e^{j\varphi(t)}\\
\end{split}
\label{eq:III.0.5}
\end{equation}
where $\Vert u(\theta,\zeta) / r \Vert$ is the amplitude scale factor induced by geometry of interfering source and the receiver. $u(t)$ is the amplitude envelope and $\varphi(t)$ the instantaneous phase of the impulsive interference. In practice, receiving the signal at the receiver is distorted due to multipath propagation. Thus, the resulting impulsive noise is a random process where amplitude envelope and instantaneous phase are random processes. The propagation law may also induce the randomness of the amplitude scale factor. In addition, a background noiseshould be considered as combination of multiple independent interference sources below impulsive interference sources \textit{e.g}, ambient noise from substations, thermal noise from receiver, etc. 


We then fully write the random process $X_{t}$ as a combination of the shot-noise process $I_{t}$ with an additive background noise $n_{t}$ such that :
\begin{equation}
X_{t} =  I_{t}+ n_{t}
   \label{eq:III.0.6}
 \end{equation}

\subsection{Non-Gaussian noise process}
\begin{figure}[!htbp]
\centering
\includegraphics[width=3.0in]{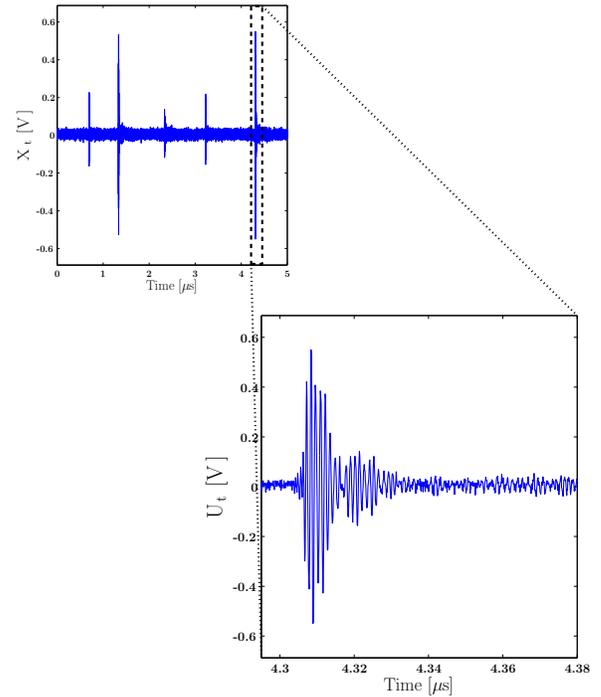}
\caption{Example impulsive noise measured in a $735$ kV substation}
\label{fig:III.A.1}
\end{figure}

A common receiver design operates at a given carrier or center frequency. Therefore, the noise process has a resonant frequency such that impulsive noise is a transient signal with damped oscillation (see Fig. \ref{fig:III.A.1}). It is seen that impulsive noise is distorted randomly due to constructive and destructive waves induced by the multipath channel related to the geometry of interference source and the receiver. It is argued that $U_{t}$ is generally non-stationary process where noise samples are non-\textit{i.i.d.} 

The receiver may have a local oscillator to recover any desired signals in baseband. In this condition, signals can be demodulated at the desired resonant frequency $\omega_{0}$. In complex domain, the baseband representation is given by :

\begin{equation}
\begin{split}
X_{t} & = \left( \vert I_{t}\vert e^{j\varphi_{I}(t)}+ \vert n_{t} \vert e^{j\varphi_{n}(t)}\right) e^{-j\omega_{0} t}\\
\end{split}
\label{eq:III.A.1}
\end{equation}
where the instantaneous phase of any analytic signal is expressed as $\varphi(t) = \sum_{i} \omega_{i} t$ where the resonant frequency $\omega_{0}$ exist in $\varphi(t)$. Baseband signals may be more tractable for impulsive noise signal processing. Indeed, the power spectrum density can be estimated by using classical parametric spectral density estimation such as Yule-Walker method. As a result, these impulsive noise can be reproduced by using discrete-time series models such as autoregressive process. 

\begin{figure}[!htbp]
\centering
\subfigure[Impulsive noise in baseband \label{A:III.A.1}]{
\includegraphics[width=1.65in]{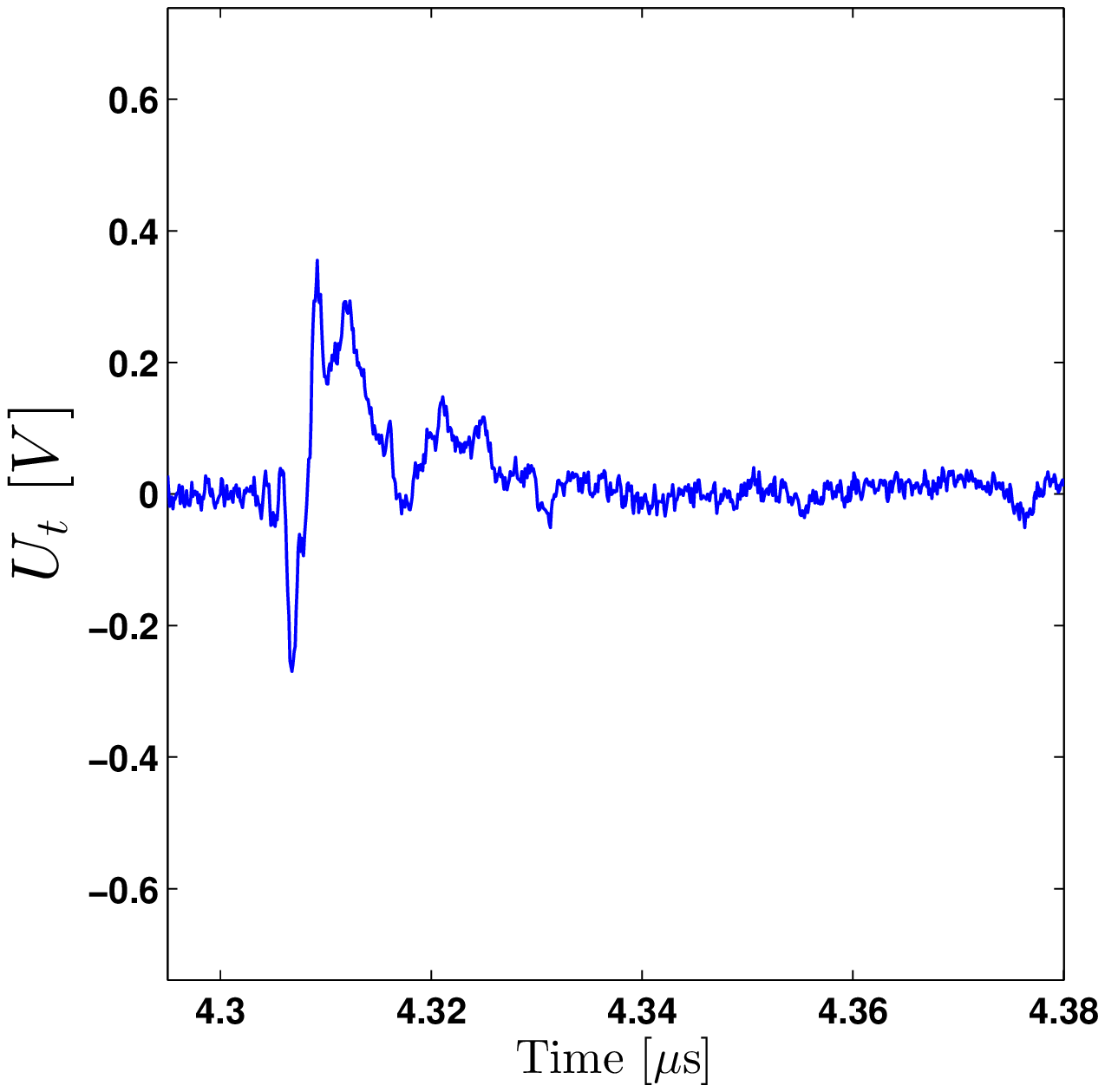}
}
\quad
\subfigure[Power spectral density \label{B:III.A.1}]{
\includegraphics[width=1.37in]{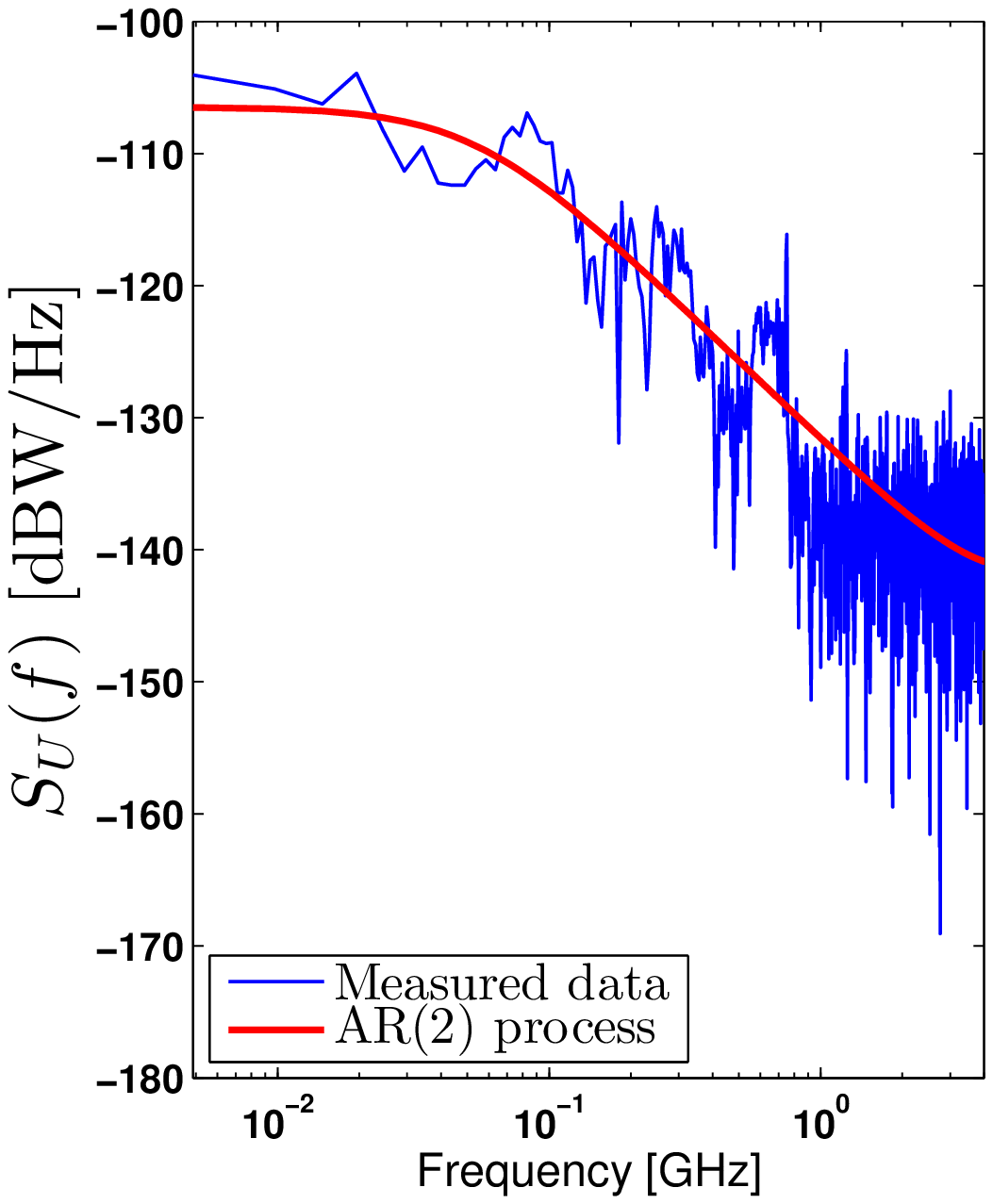}}
\caption{Impulsive noise in baseband measured in a $735$ kV substation}
\label{fig:III.A.2}
\end{figure}

Fig. \ref{fig:III.A.2} is an example of typical impulsive waveform and psd measured in a $735$ kV substation in baseband, demodulated at $f_{0} = 800$ MHz. It is seen that a second order of AR process model gives suitable estimation of the decay of $\sim 1/f^{k}$ of the noise process. The non-\textit{i.i.d} of noise samples in presence of an impulsive noise is induced by the decay of the power spectral density. The innovation process should be defined to compute distortions. The determination of first and second order statistics of the non-Gaussian process $X_{t}$ strongly depends on the specification of the impulsive shapes $U_{t}$, see the characteristic function in the equation (\ref{eq:V.C.4}). The basic waveform of the impulsive noise should take account physical parameters such as the duration of radiations, the non-stationary behaviour of the impulsive noise $U_{t}$ in which the amplitudes of the random process are non-\textit{i.i.d.}

\section{A general impulsive noise waveform model using discrete-time series}
\label{sect:IV}
In this section, a basic waveform of impulsive interference is specified. The impulse waveform at the receiver may depend on RF and IF stages at the receiver where the resulting waveform has damped oscillations generally at the carrier frequency. The general impulsive noise waveform model help to make some simple assumptions to derive first and second order statistics of the non-Gaussian process $X_{t}$. 

Discrete-time series models can compute these typical random waveforms observed from experimentations, \textit{i.e} transient impulsive noise waveform with damped oscillation, damped exponential or a mixture of damped exponential oscillation. The amplitude at the present sample denoted by $U_{t}$ depends on amplitude at the past samples denoted by $U_{t-i}$ where $i > 0$. These are weighed by coefficients which give the behaviour of the obtained waveform $U_{t}$. The definition of these coefficients should be carefully defined for the stability of the process. To make $U_{t}$ as a random process, the innovation process have to be a random variable to be defined.

\subsection{An autoregressive process for impulsive noise waveform modelling}
We consider real-valued random process $U_{t}$, the impulse shape received at the antenna produced by partial discharges as a discrete-time series such as an AR($p$) process model is given by :

\begin{equation}
\begin{split}
U_{t} &= \overset{p}{\underset{i = 1}{\sum}}\phi_{i}U_{t-i}+\varepsilon_{t}
\end{split}
\label{eq:IV.A.1}  
\end{equation}
where amplitude at the past samples $U_{t-i}$  are weighted by $\phi_{i}$ named AR($p$) coefficients. $\varepsilon_{t}$ is the innovation process that leads to distortions of $U_{t}$. We assume a second order of the AR($2$) process such that :

\begin{equation}
\begin{split}
U_{t} &= \phi_{1}U_{t-1}+\phi_{2}U_{t-2}+\varepsilon_{t}
\end{split}
\label{eq:IV.A.2}  
\end{equation}
The AR coefficients $\phi_{1}$ and $\phi_{2}$ will be defined to ensure the stability of the process, \textit{i.e} all its roots from the characteristic function lie outside the unit circle. Thus, the stationarity conditions should be verified. The second order of the AR process model allows to determined roots and the autocorrelation function of the random process $U_{t}$ easily.

\subsection{Definition of the innovation process}
The randomness of the process $U_{t}$ is induced by the innovation process $\varepsilon_{t}$. The latter should take account many random phenomena such as distortion of the impulsive shape and the non-stationary behaviour of the process linked by the duration of radiations received at the antenna. The definition of $\varepsilon_{t}$ is based on physical assumptions :

\begin{itemize}
\item When an interference source is activated, radiations received at the antenna is a superposition of the constructive and destructive impulsive waves caused by multipath effects. These can be seen as a wave distorted by an \textit{i.i.d} random variable  $\varepsilon_{t}$ such that the equation (\ref{eq:IV.A.2}) is satisfied.
\item Reflectors in substations cause multiple delayed paths that obey to the propagation law. In this condition, the amplitude of the impulse received at the antenna is necessarily decaying with respect to time until it vanishes below to the background noise \textit{i.e}, $\varepsilon_{t}$ has a time-dependent parameter denoted by $\vartheta_{t}$ with the constraint that $\varepsilon_{t}$ is a function that decay over time or samples.
\end{itemize}

From these assumption, we can define $\varepsilon_{t}$ as a white noise where the variance is a discrete-time function, \textit{i.e}, heteroscedastic white noise process \cite{Bollerslev1986,Engle1982} :

\begin{equation}
 \varepsilon_{t} = \vartheta_{t}W_{t}
 \label{eq:IV.B.1} 
 \end{equation} 
where $\vartheta_{t}$ is the time-dependent standard deviation of the white noise process $W_{t} \sim \mathcal{N}(0,1)$. For those complex-valued random process $U_{t}$, one can consider complex-valued white noise process $W_{t}$. The discrete-time function $\vartheta_{t}$ can be defined as a positive power-law or log-normal function. The latter  takes account rise time and a fall time of impulsive noise. Hence we write $\vartheta_{t}$ as :

\begin{equation}
\vartheta_{t} = \frac{\vartheta_{0}}{t\sigma_{t}\sqrt{2\pi}}\exp\left(- \frac{\left( \log t-\mu_{t}\right)^{2}}{2\sigma_{t}^{2}} \right) 
  \label{eq:IV.B.2}
\end{equation}
where $\sigma_{t}$ is related to the time decay of the impulse. $\mu_{t}$ may refer to the time where the envelope of an impulse is maximum and may be related to the presence of the main path received at the antenna. $\vartheta_{0}$ is a normalized parameter. It is convenient that these parameters should be set such that the rise time and the decay time of an impulse are much shorter than the sample size of the non-Gaussian process.  

\begin{figure}[!htbp]
\centering
\includegraphics[width=3.05in]{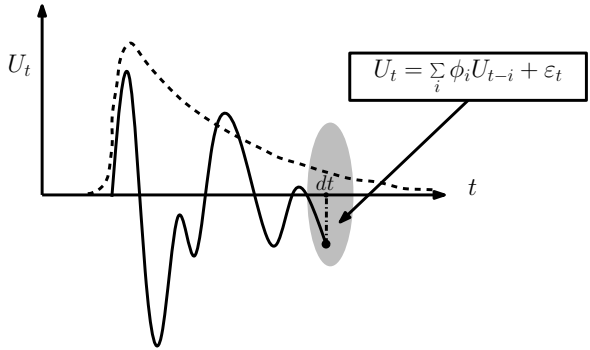}
\caption{Impulse waveform of impulsive noise distorted by a random innovation process}
\label{fig:IV.B.1}
\end{figure}

A basic waveform of impulsive noise received at the antenna has been specified based on physical assumptions. The model produce impulsive noise waveforms where amplitude are distorted randomly by the innovation process as depicted on Fig. \ref{fig:IV.B.1} where $dt$ is a time-increment defining a sample. The lined curve is the real-valued impulsive noise process where amplitudes should decay with respect to time represented by the dashed curve. The process is non-stationary due to the time-dependent of the standard deviation of the innovation process $\varepsilon_{t}$. 

\subsection{Stationarity conditions}
To ensure the stability of the process $U_{t}$, the stationarity conditions should be verified. By using the AR($2$) process model, we use the Box-Jenkins modelling approach \cite{Box1994}. The non-stationary process is differenced until stationary is achieved. Hence, from the equation (\ref{eq:IV.A.2}), we write the difference-stationary process such that :

\begin{equation}
\Phi(L)U_{t} =  \varepsilon_{t}
  \label{eq:IV.C.1}
 \end{equation} 
where $\Phi(L) = 1-\phi_{1}L-\phi_{2}L^{2}$  and $L$ is the lag operator such that $U_{t}L^{i}= U_{t-i}$. Characteristic equation of the AR($2$) process is given by :

\begin{equation}
1-\phi_{1}L-\phi_{2}L^{2} = 0 
\label{eq:IV.C.4}
\end{equation}
The quadratic equation (\ref{eq:IV.C.4}) has two roots $r_{1}$ and $r_{1}$ where :

\begin{equation}
r_{1,2} = \frac{\phi_{1}\pm \sqrt{\phi_{1}^{2}+4\phi_{2}}}{-2\phi_{2}} 
\label{eq:IV.C.4.bis}
\end{equation}
The roots depends on the value of the terms $\phi_{1}^{2}+4\phi_{2}$. The process has a stationary solution if and only if :

\begin{subnumcases}{\label{eq:IV.C.4.bis.3}}
    \phi_{2}-\phi_{1} < 1\\
     \phi_{2}+\phi_{1} < 1\\
     \vert \phi_{2} \vert < 1
\end{subnumcases}

These AR coefficients specify the behaviour of the waveform of the impulsive noise $U_{t}$. They would help us to determine the problem statement for the determination of first and second order statistics of the non-Gaussian process $X_{t}$ in section \ref{sect:V}. 
 
\subsection{Power spectral density of $U_{t}$}
By remembering that the innovation process is a heteroscedastic white noise process, the power spectral density of $U_{t}$ has a classical AR($2$) psd form given by :

\begin{equation}
S_{U}(f) = \frac{S_{\varepsilon}(f)}{\vert 1-\phi_{1}e^{j2\pi f}-\phi_{2}e^{j4\pi f}\vert^{2}}
\label{eq:IV.C.6}
 \end{equation} 
where $S_{\varepsilon}(f)$ is the psd of $\varepsilon_{t}$. By using the equation (\ref{eq:App.A.4}) in Appendix \ref{Appendix:A}, we write the complete psd of $U_{t}$ as :

\begin{equation}
S_{U}(f) = \frac{\sigma_{\vartheta}^{2}}{\vert 1-\phi_{1}e^{j2\pi f}-\phi_{2}e^{j4\pi f}\vert^{2}}
\label{eq:IV.C.8}
 \end{equation}
where $\sigma_{\vartheta}^{2}$ is the variance of $\vartheta_{t}$. The variance of the white noise is $\sigma_{W}^{2} = 1$. Depending on the roots of the AR process, the psd of $U_{t}$ has different behaviour \cite{Woodward2012} :

\begin{itemize}
\item For real roots, \textit{i.e}, the terms $\phi_{1}^{2}+4\phi_{2} \geq 0$, if the characteristic equation has at least one real roots close to the unit circle, then $S_{U}(f)$ will have peak at $f = 0$ if $\phi_{1}$ is positive. The psd will have peak at $f = 0.5$ if $\phi_{1}$ is negative. 
\item For complex roots, \textit{i.e}, the terms $\phi_{1}^{2}+4\phi_{2} < 0$, if the roots are closed to the unit circle, a peak occurs near the resonant frequency at $f_{0}$ given by :

\begin{equation}
f_{0} = \frac{1}{2\pi} \cos^{-1}\left( \frac{\phi_{1}}{2\sqrt{-\phi_{2}}}\right)
\label{eq:IV.C.9}
\end{equation}
\end{itemize}

\subsection{Autocorrelation function of $U_{t}$}
The impulse shape has a complex form due to the randomness of the amplitude. It may useful to provide the autocorrelation function (ACF) of the process $U_{t}$. From the roots of the characteristic equation on (\ref{eq:IV.C.4}), the equation (\ref{eq:IV.C.1}) is rewritten as :

\begin{equation}
U_{t} = (1-G_{1}L)^{-1}(1-G_{2}L)^{-1}\varepsilon_{t}
\label{eq:IV.C.4.bis.2}
\end{equation}
where $G_{1} = 1/r_{1}$ and $G_{2} = 1/r_{2}$ remembering that $r_{1,2}$ are the roots of the quadratic equation (\ref{eq:IV.C.4}). The autocorrelation denoted by $E\left[U_{t}U_{t-k}\right]   \equiv \rho_{k}$ of the process $U_{t}$ following a closed form solutions :

\begin{equation}
\rho_{k} = 
\begin{cases}
\frac{(1-G_{2}^{2})G_{1}^{k+1}-(1-G_{1}^{2})G_{2}^{k+1}}{(G_{1}-G_{2})(1+G_{1}G_{2})} &  \text{when $r_{1} \neq r_{2}$}\\
\\
\left( 1+\frac{(1+\phi_{2})k}{1-\phi_{2}}\right)\left(\frac{\phi_{1}}{\phi_{2}}\right)^{k}  &  \text{when $r_{1} = r_{2}$}\\
\end{cases}
\label{eq:IV.C.5} 
\end{equation} 
The behaviour of the ACF $\rho_{k}$ depends on the nature of the roots of the quadratic function : 

\begin{itemize}
\item For real roots, with the constraint that $\vert G_{1} \vert$ and $\vert G_{2}\vert <1$, the ACF $\rho_{k}$ can be seen as mixture of damped exponentials or damped exponential oscillation that decay to zeros when $k$ increases. 
\item For complex roots, the ACF is a damped sinusoidal function where the explicit expression of $\rho_{k}$ is given by 
\cite{Woodward2012} :

\begin{equation}
\left(\sqrt{-\phi_{2}}\right)^{k}  \frac{\sin\left(2\pi k f_{0} + \varsigma \right) }{\sin \left( \varsigma \right)  } 
\label{eq:IV.C.5.biss}
\end{equation}
where $f_{0}$ is the resonant frequency of the system and $\varsigma = (1-\phi_{2})/(1+\phi_{2}) \cdot \tan (2\pi f_{0})$.  
\end{itemize}

A general impulsive noise $U_{t}$ has been specified by using AR process model. If the stationarity condition of the process is ensured, one can reproduce a complete random waveform impulsive noise $U_{t}$ with damped oscillation at a desired resonant frequency by set the AR coefficient $\phi_{1}$ and $\phi_{2}$ such that the roots of the characteristic equation has complex roots. The impulsive noise can be represented in baseband where the psd $S_{U}(f)$ should have a peak at $f = 0$. In this condition, the roots of the characteristic equation have to be real and particularly the first AR coefficient has to be positive, $\phi_{1}> 0$.

We are now ready to derive the first and second order statistics of the non-Gaussian noise process $X_{t}$ based on the equation (\ref{eq:V.C.4}) by using the basic waveform $U_{t}$ of an impulsive interference from AR process model. However, it may be a non-trivial task since the impulsive noise shape has random amplitude. Some simplifying assumptions have to be considered. The second order statistics is the power spectrum of the process. It is given by the Carson's theorem \cite{Carson1931,Rice1944}.

\section{First and second order statistics of the non-Gaussian process}
\label{sect:V}

First and second order statistics are the first interest for implementation of threshold algorithms for signal detection and estimation. It may be difficult to provide exact analytical probability density function and power spectral densities when those results depend on the impulsive waveform of interference sources. In presence of impulsive interference sources in substation environments, many random impulses can be observed for a given time observation where inter-arrival time, energy of individual impulsive noise, and occurrences are randomly distributed.

\subsection{Problem statement}
We need to made some simplifying assumptions in terms of statistics from the basic Poisson field of interferers as well as in terms of basic waveforms of impulsive noise. 

\subsubsection{The homogeneity of the Poisson field of interferers}

Interference sources are generally detected in presence of HV equipments under voltage and the Poisson process $N_{\mathbf{r}}$ in space domain can be homogeneous. However, most impulsive interferences are generated by AC voltages in substation environments. Thus, the interfering sources are activated whenever the electric field reaches the dielectric strength of the air. Hence, the Poisson process $N_{t}$ in time domain may be inhomogeneous and cyclostationary. However, the presence of the three phase voltages the superposition of the activated sources may become homogeneous Poisson process. In this condition, we assume the presence of interference sources driven by the three phase of voltages independently. Therefore, a constant density of the Poisson process $\lambda(\psi) = \lambda$ is assumed. 

Furthermore, we assume a large time observation to have a non-negligible number of impulsive noise. As a result, we may write the non-Gaussian noise process $X_{t}$ as a superposition of shot-noise process $I_{t}$ produced by each individual interference source plus an additive background noise $n_{t}$ such that :

\begin{equation}
\begin{split}
X_{t} & = \sum\limits_{j = 1}^{N_{e}}  K_{\mathbf{r}_{j}}I_{t\vert  d(\mathbf{r}_{j},\mathbf{r}_{0})} + n_{t}\\
\end{split}
\label{eq:V.B.1}
\end{equation}
where $K_{\mathbf{r}_{j}}$ is a random amplitude scale factor induced by the geometry of individual interference source and the antenna. $N_{e}$ is the number of the interference sources activated in the vicinity of the antenna.

\subsubsection{The basic waveform of impulsive noise $U_{t}$}

First order statistics can be difficult to derived especially when $U_{t}$ is a random process. Nevertheless, a suitable approximation can be obtained by finding an equivalent deterministic of these impulse shapes. In \cite{Gilbert1960,Lowen1990} suggest that the equivalent function of $U_{t}$ can be determined by using the expected value $E\left[U_{t}\right]$ denoted by $\gamma_{t}$. From the equation (\ref{eq:IV.C.1}), we write the equivalent impulsive noise function as :

\begin{equation}
\Phi(L)\gamma_{t} =  E\left[ \varepsilon_{t}\right] 
  \label{eq:IV.C.2}
 \end{equation} 
The innovation process is a heteroscedastic white noise. Therefore, $E\left[ \varepsilon_{t}\right] = 0$. Thus, the expected value of $U_{t}$ is derived from the second order of the difference equation (\ref{eq:IV.C.2}). We see that the solution of this equation depends on the roots of the characteristic function. 

\begin{itemize}
\item For real roots, a solution of the equation (\ref{eq:IV.C.2}) is :
\begin{equation}
 \gamma_{t} = K \left(e^{-a t}- e^{-b t}\right) 
  \label{eq:IV.C.3}
 \end{equation} 
where $b$ is related to the rise time, $a$ is the fall time of the discharge $\gamma_{t}$ and $K$ is a random amplitude factor.

\item For complex roots, a solution of the equation (\ref{eq:IV.C.2}) is written as : 
\begin{equation}
 \gamma_{t} = K \left(e^{-a t}- e^{-b t}\right)\cos(2\pi f_{0}t+\varphi)
  \label{eq:IV.C.3.bis}
 \end{equation}
where the resonant frequency of the system is given the equation (\ref{eq:IV.C.9}) and $\varphi$ an arbitrary phase. Here, $b$ and $a$ is not necessarily equal as find on the equation (\ref{eq:IV.C.3}).
\end{itemize}

It is convenient for impulsive waveforms that $a$ and $b$ are real numbers strictly positive. Moreover, we restrict the mathematical development of first and second order statistics for baseband impulsive noise. Nevertheless, one can follow the same approach if the damped oscillation in the equation (\ref{eq:IV.C.3.bis}) need to be considered. Therefore, first and second order statistics are derived based on the waveform from the equation (\ref{eq:IV.C.3}). One can recover impulsive noise with damped oscillations by multiplying the impulse in baseband by a carrier wave at the desired resonant frequency $f_{0}$. We consider the basic waveform $\gamma_{t}$ as a continuous-time function and we assume that the shot-noise process $I_{t}$ and the background noise $n_{t}$ are independent random processes. 

\subsection{Moments and cumulants}
The description of the shape of amplitude distributions and densities can be given, in some extent, by moments and cumulants. For example, the skewness is a measure the asymmetry of noise and the kurtosis, a measure of how outlier-prone the distribution is. We start by calculating the $m^{\mbox{\small{th}}}$ cumulant $\kappa_{m}$ of the shot-noise process $I_{t}$. In equation (\ref{eq:IV.C.3}), we assume $K$ as a random variable which assume positive and negative values make distributions. Then, from the extension of the Campbell's theorem, the $m^{\mbox{\small{th}}}$ cumulant $\kappa_{m}$ is given by \cite{Rice1944} :

\begin{equation}
\begin{split}
\kappa_{m} & \equiv \frac{\partial^{m} }{\partial s^{m}} \log\left[  Q_{I}(s) \right]   \bigg\vert_{s = 0}\\
& = \lambda \left\langle   \int_{\mathcal{R}} \gamma_{t}^{m} dt \right\rangle  \\
& = \lambda \left\langle  K^{m}\right\rangle  \int_{\mathcal{R^{+}}} \left( e^{-a t}-e^{-b t}\right)^{m} dt 
\end{split}
\label{eq:V.B.2}
\end{equation}
where $Q_{I}(s)$ is the first order moment generating function of the shot-noise process. $\left\langle \cdot \right\rangle $ is the expectation taken over the distribution of the random variable. From binomial formula, we write the $m^{\mbox{\small{th}}}$ cumulant $\kappa_{m}$ as :
\begin{equation}
\begin{split}
\kappa_{m} & = \lambda \left\langle  K^{m}\right\rangle \int_{\mathcal{R^{+}}} \sum\limits_{k = 0}^{m}{m \choose k} \left(e^{-a t} \right)^{m-k} \left(-e^{-b t} \right)^{k} dt\\
 & =  \lambda \sum\limits_{k = 0}^{m}{m \choose k}(-1)^{k} \frac{\left\langle K^{m}\right\rangle }{a(m-k)+b k}\\
\end{split}
\label{eq:V.B.3}
\end{equation}
A series can be identified with a binomial sequence. By using the ratio test, we prove that $\kappa_{m}$ is convergent for infinite value of $m$ as seen in appendix \ref{Appendix:B}. Hence, $\kappa_{m}$ is necessarily finite. We extend to the non-Gaussian noise and by assuming the independence between the shot-noise process and the background Gaussian noise. Hence, we have :
\begin{equation}
\begin{split}
\kappa_{m} & \equiv \frac{\partial^{m} }{\partial s^{m}} \log\left[  Q_{X}(s) \right]   \bigg\vert_{s = 0}\\
& =  \frac{\partial^{m} }{\partial s^{m}} \log\left[  Q_{I}(s) \right] + \frac{\partial^{m} }{\partial s^{m}} \log\left[  Q_{n}(s) \right]   \bigg\vert_{s = 0} 
\end{split}
\label{eq:V.B.4}
\end{equation}
Finally, for an additive background noise in which an \textit{i.i.d} Gaussian noise of zero mean and variance $\sigma_{n}^{2}$ is assumed, $n_{t} \sim \mathcal{N}(0,\sigma_{n}^{2})$. Thus, we have :

\begin{subequations}
\begin{align}
\kappa_{1} & =  \lambda \frac{ \left\langle  K \right\rangle (b-a) }{ab}\\
\kappa_{2} & =  \lambda \sum\limits_{k = 0}^{2}{2 \choose k}(-1)^{k}\frac{ \left\langle  K^{2}\right\rangle }{a(2-k)+b k}+ \sigma_{n}^{2}\\
\kappa_{m} & = \lambda \sum\limits_{k = 0}^{m}{m \choose k}(-1)^{k}\frac{\left\langle  K^{m}\right\rangle }{a(m-k)+b k}  \bigg\vert_{m > 2}
\end{align}
\label{eq:V.B.5}
\end{subequations}
We assume a non-negative impulsive shape such that $b$ higher than $a$. In this condition, for a finite value of $m = \lbrace 1, 2, 3, 4, 5, 6\rbrace$, the $m$-cumulants $\kappa_{m}$ is a non-monotonic sequence if $\left\langle  K^{m} \right\rangle $ is negative for odd values of $m$, \textit{i.e} $\kappa_{m}$ is negative for odd values and $\kappa_{m}$ is positive otherwise. The $m$-cumulants can be linked with the $m$-moments of the non-gaussian model. The skewness $\chi_{1}$ and the kurtosis $\chi_{2}$ are determined by the  $3^{\mbox{\small{rd}}}$ and the $4^{\mbox{\small{th}}}$ standardized moments of the non-Gaussian noise $X_{t}$ respectively. They can be written in terms of cumulants such that : 

\begin{subequations}
\begin{align}
\chi_{1} & = \frac{\kappa_{3}}{\kappa_{2}^{3/2}} \\
\chi_{2} & = \frac{\kappa_{4}}{\kappa_{2}^{2}}
\end{align}
\label{eq:V.B.6}
\end{subequations}

\begin{itemize}
\item The skewness value only depends on $\left\langle  K^{3}\right\rangle $. Indeed, if the skewness is $\vert \chi_{1} \vert  \leq 0$, then the probability density function of amplitude (pdf) can be \textit{left-skewed} or \textit{right-skewed}, \textit{i.e} presence of longer tail on the left or on the right. The pdf can be also symmetric if $\left\langle  K^{3}\right\rangle = 0$. In practice, the asymmetry may be induced by random distortions of impulse shapes.
\item In presence of impulsive noise, if $\left\langle K^{4}\right\rangle >0 $ and $\kappa_{4}>\kappa_{2}^{2}$, then the kurtosis is always $\chi_{2} >0 $. Hence, the pdf can be \textit{leptokurtic}, \textit{i.e} a peak around the mean and long tail at higher amplitude values.
\end{itemize}

\subsection{Moment generating function and characteristic function}

The moment generating function of the non-Gaussian noise is expressed by :

\begin{equation}
\begin{split}
Q_{X}(s) & \equiv E\left[ e^{-sX}\right]\\
& = E\left[ e^{-sI} e^{-s  n}\right]\\
& = Q_{I}(s) Q_{n}(s) \\
\label{eq:V.B.7}
\end{split}
\end{equation}
where $s\in \mathcal{R}$ and $Q_{I}(s)$ and $Q_{n}(s)$ are the moment generating function of the shot-noise $I_{t}$ and the additive background Gaussian noise $n_{t}$. The moment generating function of Gaussian noise $Q_{n}(s)$ is given by :

\begin{equation}
Q_{n}(s) = \exp\left( \frac{\sigma_{n}^{2}s^{2}}{2}\right) 
\label{eq:V.B.8}
\end{equation}

The generating function of the shot-noise process is more difficult to obtain. Nevertheless, a closed expression form of the moment generating function can be provided in terms of cumulant and by using the series expansion. We emphasize that the cumulant $\kappa_{m}$ is finite when $m$ goes to infinite as we proved in Appendix \ref{Appendix:B}. Hence, we start by the series expansion of the cumulant generating function of the shot-noise process such that :

\begin{equation}
\begin{split}
\log \left[ Q_{I}(s)\right]  & = \sum\limits_{m = 1}^{\infty} \frac{\kappa_{m}s^{m}}{m !}\\
& = \lambda \sum\limits_{m = 1}^{\infty} \sum\limits_{k = 0}^{m}{m \choose k}\frac{(-1)^{k} s^{m}}{m !}\frac{\left\langle  K^{m}\right\rangle }{a(m-k)+b k}
\end{split}
\label{eq:V.B.9}
\end{equation}
Extended to the non-Gaussian noise, we link all values of the cumulant $\kappa_{m}$ as defined in equations (\ref{eq:V.B.5}) to the moments of the distribution. The cumulant generating function can be linked by the moment generating function $Q_{X}(s)$ where the cumulant generating function is the logarithm of the moment generating function :

\begin{equation}
\begin{split}
Q_{X}(s) & = 1+\sum\limits_{m = 1}^{\infty} \frac{\mu_{m} s^{m}}{m !}\\ 
 & = \exp\left( \sum\limits_{m = 1}^{\infty} \frac{\kappa_{m}s^{m}}{m !} \right)
\end{split}
\label{eq:V.B.10}
\end{equation}
where $\mu_{m}$ are moments of the distribution of the non-Gaussian noise such that : 

\begin{subequations}
\begin{align}
E\left[X_{t} \right] = \mu_{1} & =  \kappa_{1}\\
E\left[X_{t}^{2} \right] = \mu_{2} & =  \kappa_{2}+\mu_{1} \kappa_{1}\\
E\left[X_{t}^{3} \right] = \mu_{3} & = \kappa_{3}+2\kappa_{2}\mu_{1}+\kappa_{1}\mu_{2}\\
\cdots
\end{align}
\label{eq:V.B.11}
\end{subequations}
The cumulant generating function can be seen as an entire function for complex values of $s$ where the function converges everywhere in the complex plane, (see Appendix \ref{Appendix:B}). Thus, the characteristic function $Q_{X}(j\xi)$ can be expressed by replacing $s = j\xi$ in the equation (\ref{eq:V.B.10}). 

\subsection{Amplitude probability distribution  and density of non-Gaussian noise process}
\subsubsection{General expression of pdf}
The general expression of probability density function of amplitude of $X_{t}$ can be given by the convolution product of the shot-noise process $I_{t}$ and the background Gaussian noise $n_{t}$ due to the independence of $I_{t}$ and $n_{t}$. We write the pdf $f_{X}(x)$ as :

\begin{equation}
\begin{split}
 f_{X}(x) & = f_{I+n}(x)\\
  & = \int_{\mathcal{R}} f_{I}(u)f_{n}(x-u)du\\
  & = \int_{\mathcal{R}} Q_{I}(j\xi)Q_{n}(j\xi)e^{-j\xi x}d\xi
\end{split}
\label{eq:V.B.12}
 \end{equation} 
The pdf and the characteristic function of the background Gaussian noise $n_{t}$ is well known but pdf of the shot-noise process $I_{t}$ has complex form. A general form of the pdf can be obtained by the inverse of the Fourier transform of the characteristic function of the shot-noise process \cite{Rice1944,Lowen1990} such that :

\begin{equation}
f_{I}(x) = \frac{1}{2 \pi} \Re\left\lbrace \int_{\mathcal{R}} \exp \left(-\lambda\int_{\mathcal{R}}\left[1-e^{j\xi \gamma_{t}} \right]dt \right)e^{-j\xi x} d\xi \right\rbrace 
\label{eq:V.B.13}
\end{equation}

Extended to the non-Gaussian noise, and by using the characteristic function of $Q_{X}(j\xi)$ based on the equation (\ref{eq:V.B.10}), we have :

\begin{equation}
f_{X}(x) =  \frac{1}{2 \pi} \Re\left\lbrace \int_{\mathcal{R}} \exp \left( \sum\limits_{m = 1}^{\infty} \frac{\kappa_{m}(j\xi)^{m}}{m !}-j\xi x \right) d\xi \right\rbrace 
\label{eq:V.B.14}
\end{equation}
The pdf $f_{X}(x)$ should be bounded in $\left[0,1 \right] \forall x \in \mathcal{R}$. We may rewrite the pdf by setting :

\begin{subequations}
\begin{align}
\sigma^{2} &= \kappa_{2}\\
\nu &= \frac{x-\kappa_{1}}{\sigma}
\end{align}
\label{eq:V.B.14.bis}
\end{subequations}
From the equation (\ref{eq:V.B.14}), we rewrite the pdf as :

\begin{equation}
f_{X}(x) =  \frac{1}{2 \pi} \Re\left\lbrace \int_{\mathcal{R}} e^{ -j\xi \sigma\nu-\frac{\xi^{2}\sigma^{2}}{2}} \exp \left( \sum\limits_{m = 3}^{\infty} \frac{\kappa_{m}(j\xi)^{m}}{m !} \right) d\xi \right\rbrace 
\label{eq:V.B.14.bis.1}
\end{equation}
The expression of the pdf is complex to derived. An approximation approach can help to achieve tractable forms. 

\subsubsection{Series approximation of pdf}
The complex form of pdf in equation (\ref{eq:V.B.14}) may be approximated by series approximation \cite{Rice1944,Kolassa2006}. From the equation (\ref{eq:V.B.13}) it is convenient to define :

\begin{equation}
H(j\xi) = \frac{1}{T}\int_{-T/2}^{T/2} e^{j\xi\gamma_{t}}dt
\label{eq:V.B.14.bis.bis}
 \end{equation} 
where $H(j\xi)$ is seen as the characteristic function of $\gamma_{t}$. In this condition, the pdf of the shot-noise process in the equation (\ref{eq:V.B.13}) can be written as :

\begin{equation}
f_{I}(x) = \frac{1}{2 \pi} \Re\left\lbrace \int_{\mathcal{R}} \exp \left(\lambda T H(j\xi)-\lambda T -j\xi x \right)d\xi \right\rbrace 
\label{eq:V.B.14.bis.bis.1}
\end{equation}
Thus, from the equation (\ref{eq:V.B.14.bis.1}), we consider that :

\begin{equation}
\begin{split}
\frac{1}{2 \pi} \int_{\mathcal{R}} (j\xi\sigma)^{m}e^{ -j\xi \sigma\nu-\frac{\xi^{2}\sigma^{2}}{2}} d\xi & = (-1)^{m}\sigma^{-1}\Theta^{m}(\nu)
\end{split}
\label{eq:V.B.14.bis.2}
\end{equation}
where :

\begin{equation}
\begin{split}
 \Theta^{(m)}(\nu) &= (2\pi)^{-1/2} \frac{\partial^{m}}{\partial \nu^{m}} e^{-\nu^{2}/2}
\end{split}
\label{eq:V.B.14.bis.2.1}
\end{equation}
Hence, the pdf of instantaneous amplitude of $X_{t}$ is given asymptotically by collecting terms according to power of $\lambda^{-1/2}$ \cite{Rice1944,Wallace1958,Middleton1974} :

\begin{multline}
f_{X}(x) \approx   e^{-\lambda T}\sum\limits_{m}^{\infty}\frac{(\lambda T)^{m}}{m !}\left\lbrace \sigma^{-1}\Theta^{(0)}(\nu)-\frac{\kappa_{3}\sigma^{-4}}{3!}\Theta^{(3)}(\nu) \right.\\\ 
+\left[\frac{\kappa_{4}\sigma^{-5}}{4!}\Theta^{(4)}(\nu)+ \frac{\kappa_{3}^{2}\sigma^{-7}}{72}\Theta^{(6)}(\nu)\right]+ \cdots \bigg\rbrace 
\label{eq:V.B.14.bis.3}
\end{multline}
where the first term is $o(\lambda^{-1/2})$ which is the normal distribution, the second term is  $o(\lambda^{-1})$ and terms within brackets is $o(\lambda^{-3/2})$. The approximation is based on the Edgeworth series. By considering only the first term and linking the standard deviation $\sigma$ with the increment $m$ such that $\sigma_{m} = g(\sigma,m)$ a function of $\sigma$ and $m$ in the equation (\ref{eq:V.B.14.bis.3}), one can find the Middleton Class A \cite{Middleton1974} such that :

\begin{equation}
f_{X}(x) \approx   e^{-\lambda T}\sum\limits_{m}^{\infty}\frac{(\lambda T)^{m}}{m !} e^{-x^{2}/2\sigma_{m}^{2}}
\label{eq:V.B.14.bis.bis.3}
\end{equation}
However, the Edgeworth series expansion is often inaccurate in the far tail of distribution \cite{Daniels1987,Lugannani1980}.

\subsubsection{Convergence to $\alpha$-stable distribution}
The non-Gaussian noise process can be seen as a sum of independent processes where the shot noise $I_{t}$ is written as a sum of independent processes $U_{t}$ such that :

\begin{equation}
X_{t} = \frac{U_{t,1}+U_{t,2}+\cdots+U_{t,m}}{d_{m}}+n_{t}
\label{eq:V.B.14.bis.4}
\end{equation}
where $d_{m}$ is a sequence of positive real numbers strictly positive. By definition, the process $X_{t}$ is stable \cite{Samarodnitsky2000}. The random process $U_{t}$ is impulsive noise where its distribution is $f_{\vert U \vert}(u) \sim \vert u\vert^{-\alpha-1}$ where $\alpha$ is the characteristic exponent. From \cite{Lowen1989,Gilbert1960}, in absolute values, one can find that the distribution of basic waveforms write in equations (\ref{eq:IV.C.3}) and (\ref{eq:IV.C.3.bis}) are $ \sim \vert \gamma\vert^{-\alpha-1}$ where $0< \alpha < 2$. As a result, the random process $X_{t}$ has a $\alpha$-stable distribution such that the characteristic function is \cite{Gnedenko1954,Samarodnitsky2000} :

\begin{equation}
\begin{split}
Q_{X}(j\xi) & =  \exp\left\lbrace j\xi \mu -\vert\sigma \xi\vert^{\alpha}\left( 1-j\beta\operatorname{sign}(\xi)\eta \right)  \right\rbrace 
\end{split}
\label{eq:V.B.14.bis.5}
\end{equation}
where $\mu$ is a location parameter real value, $\sigma \geq 0$ is a scale factor, $\beta$ is the skewness parameter where $-1 \leq \beta \leq 1$, $\eta = \tan(\pi\alpha/2)$ if $\alpha \neq 1$ and $\eta = \log\vert\xi\vert$ if $\alpha = 1$. 
  
We emphasize that the pdf of the non-Gaussian noise is a fat-tailed distribution with high value of kurtosis and it can be also asymmetric as argued. The energy and the duration of impulsive noise determine, to some extent, the ``fatness" of the tail of the distribution. These parameters increase the probability of amplitude values higher than its standard deviation. The Midlleton Class A can approach the non-Gaussian noise however, it may be inaccurate in the tail of distribution. The $\alpha$-stable can provide a suitable approximation of the amplitude distribution for those random processes which admit a power law decay of $\sim \vert x\vert^{-\alpha-1}$ on distribution. The two approximations will be compared in the section \ref{sect:VI}. The tail distribution is given by :

\begin{equation}
\bar{F}_{X}(x) = P(X>x) = 1-\int_{-\infty}^{x}f_{X}(u)du
\label{eq:V.B.14.bis.6}
\end{equation}

\subsection{Second order statistics : Power spectral density of $X_{t}$}
\label{sect:VI.D}
The power spectral density of the shot-noise process $I_{t}$ can be given in terms of the rate $\lambda$ and the Fourier transform of the impulse response of the associated linear filter by the Carson's theorem \cite{Carson1931,Rice1944}. The power spectral density is given by the autocorrelation function of the non-Gaussian noise process $X_{t}$ is given by \cite{Rice1944,Rice1945,Lowen1990} :

\begin{equation}
E\left[X_{t}X_{t+\tau} \right]  = E\left[I_{t}^{2}\right]+\lambda E\left[\gamma_{t}\gamma_{t+\tau}\right]+\sigma_{n}^{2}\delta(\tau)
\label{eq:V.B.15.bis}
\end{equation}

The Carson's theorem allow to express the power spectral density with these terms if the integral of the autocorrelation function of $\gamma_{t}$ is finite or equivalently, the integral of the psd is finite \cite{Rice1944,Lowen1990}. We assume that $\left\langle K^{2}\right\rangle < \infty $. One can find the integral of psd of $\gamma_{t}$ is finite, $\int_{\mathcal{R}} S_{\gamma}(f) df < \infty ~\forall f\in \mathcal{R}$. In this condition, extended to the non-Gaussian process, the resulting expression of the psd of $X_{t}$ is given by :

\begin{equation}
S_{X}(f) = E\left[ I_{t}^{2}\right]\delta(f)+\frac{\lambda  \left\langle K^{2} \right\rangle (b-a)^{2}}{(a^{2}+\omega^{2})(b^{2}+\omega^{2})}+\sigma_{n}^{2}
\label{eq:V.B.15}
\end{equation}
where $\omega = 2\pi f$. It is seen that the psd of the non-Gaussian noise process has a decay of $\sim 1/f^{k}$.

In this section, the first and second order statistics can be derived from the basic waveform of the impulsive noise. We proved that we can have high value of kurtosis in which the distribution is \textit{leptokurtic} and also be asymmetric as discussed. Furthermore, we proved that amplitude distributions and densities of the non-Gaussian process can be approximated by classical non-Gaussian pdf forms such as Middleton class A or $\alpha$-stable. The power spectral density can also be derived where a decay of $\sim 1/f^{k}$ is observed induced by waveforms of impulsive noise.

\section{Results and discussion}
\label{sect:VI}
In this section, we discuss about the validity of our theoretical model when the electromagnetic environment has impulsive interferers where the resulting waveforms at a receiving point is a succession of independent impulsive noise. 

We will start by specifying waveforms of impulsive noise using discrete-time series. We will define coefficients in which the stationarity condition is ensured for the random impulsive waveform process $U_{t}$, see equation (\ref{eq:IV.C.4.bis.3}). Thus, a non-Gaussian noise process $X_{t}$ can be fully simulated where a succession of random impulsive noise is excited by Poisson point process. Additionally, a background noise $n_{t}$ below the shot-noise process $I_{t}$ is considered. In this condition, the first order can be derived where empirical amplitude distribution and density can be provided. We show how classical non-Gaussian noise model such as Midlleton class A and $\alpha$-stable amplitude distribution and density can be appropriated vis-a-vis the simulation results as well as vis-a-vis real situations in substation environments. 

\subsection{Impulsive waveforms modelling}

\begin{figure}[!htbp]
\centering
\subfigure[Amplitude \label{A:VI.A.1}]{
\includegraphics[width=1.57in]{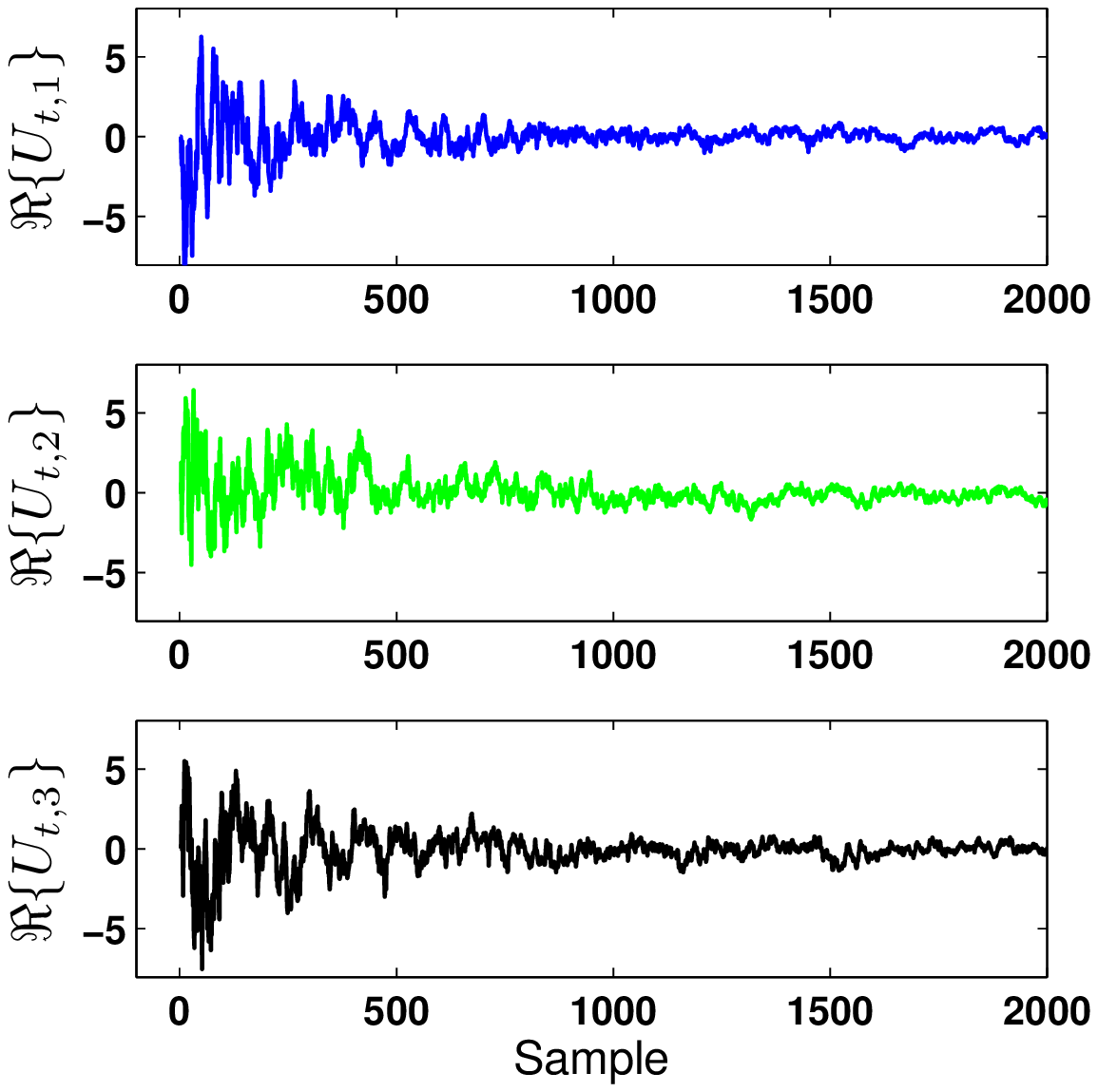}
}
\quad
\subfigure[Envelope \label{B:VI.A.1}]{
\includegraphics[width=1.57in]{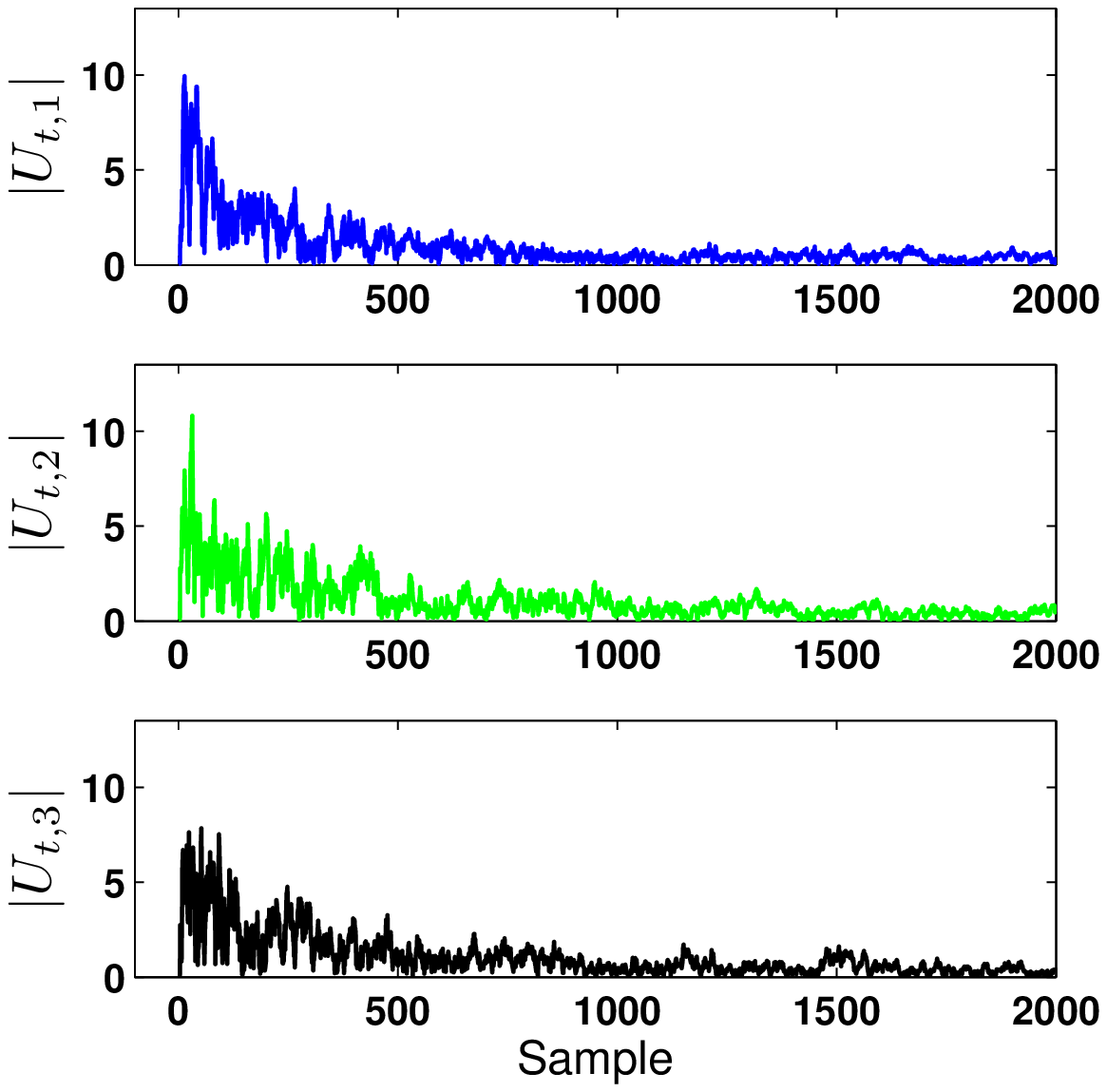}}
\quad
\subfigure[Power spectral densities \label{C:VI.A.1}]{
\includegraphics[width=1.55in]{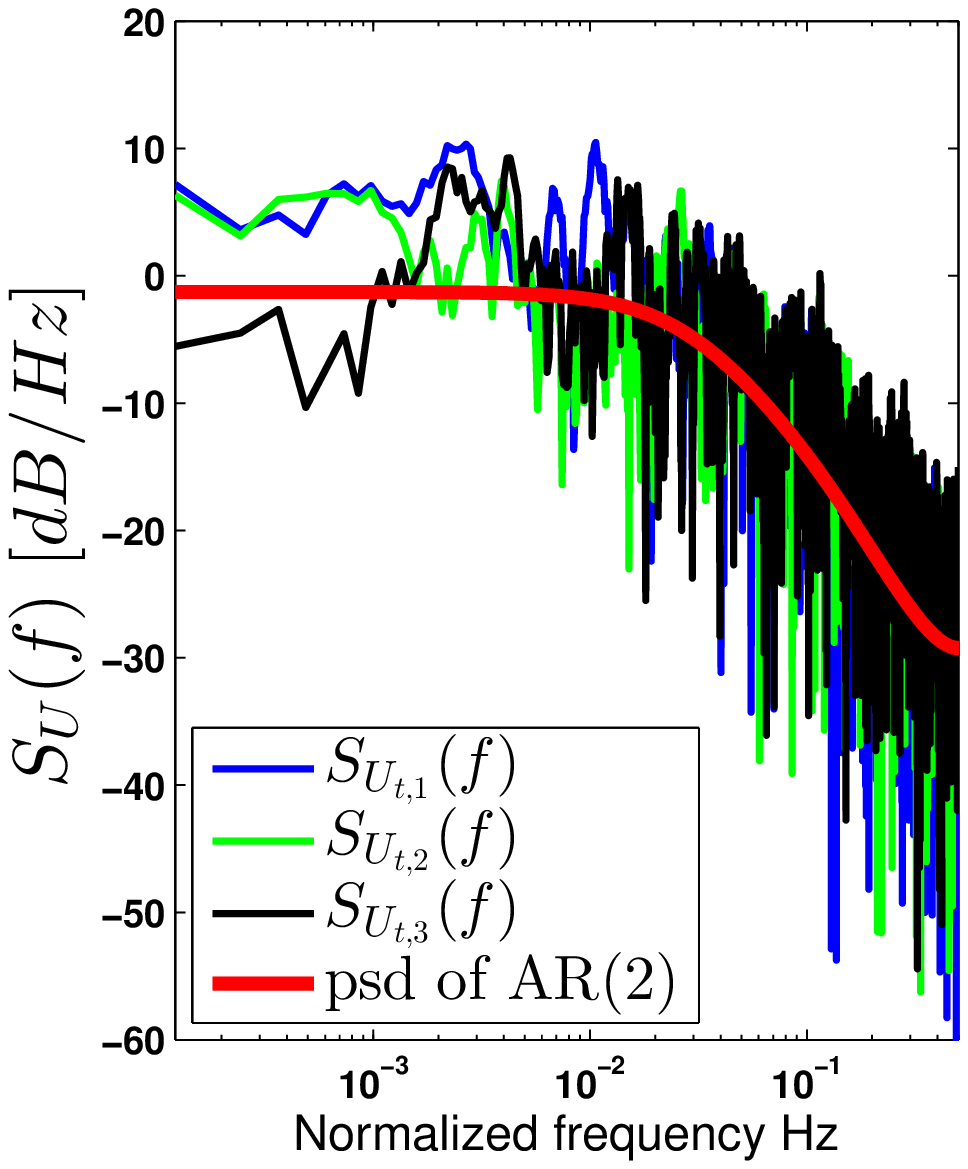}}
\caption{Example of random impulsive noise $U_{t}$}
\label{fig:VI.A.1}
\end{figure}

In section \ref{sect:IV}, a complete random impulsive noise can be computed based on discrete-time series such as AR process. According to the equation (\ref{eq:IV.A.2}), a second order of the AR process is used. We only restrict the discussion where impulsive noise are in baseband, \textit{i.e}, a decay of $\sim 1/f^{k}$ with a peak at $f = 0$. In this condition, we restrict AR coefficients such that their roots is real values and the stationarity condition is ensured, see equations (\ref{eq:IV.C.4}) and (\ref{eq:IV.C.4.bis.3}). The innovation process induced by $\varepsilon_{t}$ is a heteroscedastic white noise process where the time-dependent standard deviation is given by the equation (\ref{eq:IV.B.2}). Parameters of the latter must be set such that the rise time and the decay time of a random impulsive waveform $U_{t}$ are much shorter than the sample size of the non-Gaussian process. Many random impulsive noise can be simulated as depicted on Fig. \ref{fig:VI.A.1} where parameters are set as follows :

\begin{table}[!htbp]
\caption{Impulsive noise shape parameters}
\centering
\renewcommand{\arraystretch}{1.5}
\begin{tabular}{|c|c||c|c|}
\hline
\multicolumn{2}{ |c|| }{AR coeff. $\phi$}&\multicolumn{2}{ |c| }{$\vartheta_{t}$ std. of $\varepsilon_{t}$}\\
\hline
$\phi_{1}$ & $\phi_{2}$ & $\mu_{t}$ & $\sigma_{t}$\\
\hline
$1.2$ & $-0.3$ & $7.0$ & $2.25$\\
\hline 
\end{tabular} 
\label{table:VI.A.1}
\end{table}

We see that the random process can generate many impulsive noise with random amplitudes but the power spectral densities have same behaviour, \textit{i.e} a decay of $\sim 1/f^{k}$ closed to a Lorentzian form. The desired rise time and the decay time defining the duration of the impulse can be defined by setting parameters of the time-dependent standard deviation of the heteroscedastic process. One can be able to compute many random impulsive waveforms with many behaviours based on AR process as long as the stationarity condition is ensured.

\subsection{First and second order statistics of non-Gaussian noise process}
Computer simulations and measurements are provided to validate the analysis. 
  
\subsubsection{Simulation setup}
We are now ready to simulate a non-Gaussian noise process $X_{t}$ in presence of non negligible number of impulsive noise. First of all, the electromagnetic environment should be specified by assuming : 
 
\begin{itemize}
\item A homogeneous random space-time Poisson field of interference sources where the density is an arbitrary constant positive value $\lambda(\psi) =  \lambda < 1$. 
\item All radiations from interference sources emit impulsive noise such that parameters set on the table \ref{table:VI.A.1} are satisfied.
\item The energy denoted by $\Vert U_{t} \Vert^{2} $ of each impulse is randomly distributed induced by charges and currents of partial discharge sources \cite{Brunt1991,Schifani1999}. We choose an exponential law where the energy in average, denoted by $\left\langle \Vert U_{t} \Vert^{2} \right\rangle $, is above the background noise such that the variance ratio between the background noise the shot-noise process is :

\begin{equation}
\Gamma = \frac{E\left[ n_{t}^{2}\right]}{E\left[ I_{t}^{2}\right]} < 1
\label{eq:VI.A.1.bis}
 \end{equation}  
\end{itemize}

In this condition, we set parameters as follows : the density of the Poisson field interferers $\lambda(\psi) $ is homogeneous and constant and set to $\lambda = \lambda_{t}\lambda_{\mathbf{r}}$ where the average interference sources is $\lambda_{\mathbf{r}} = 5$ per unit volume and the average radiation emissions per source is $\lambda_{t} = 5$ per sample or per unit time. The energy is random variable exponentially distributed where the average value is $\left\langle \Vert U_{t} \Vert^{2} \right\rangle = 10$. The variance ratio between the background noise the shot-noise process is $\Gamma = 0.1$.

\subsubsection{Measurement setup}
Measurement campaign is made in a $735$ kV substation. The measurement setup includes a wideband antenna ($0.8$ to $3$ GHz), RF and IF stages such as high pass filter, amplifier, etc. For data acquisition, we use an oscilloscope to capture waveforms in presence of impulsive noise. The sample rate is $10$ Gs/s for an observation time at $5~\mu$s. Details about parameters of the environment during the measurement campaign and the measurement setup is given in \cite{Au2013}. The obtained waveforms contain background noise including wireless communications from cellular or communication in the ISM band. We demodulate the received signals at the resonant frequency, $f_{0} = 800$ MHz, to obtain waveforms in baseband.

\begin{figure}[!htbp]
\centering
\subfigure[Amplitude Computer simulation \label{A:VI.B.1}]{
\includegraphics[width=1.57in]{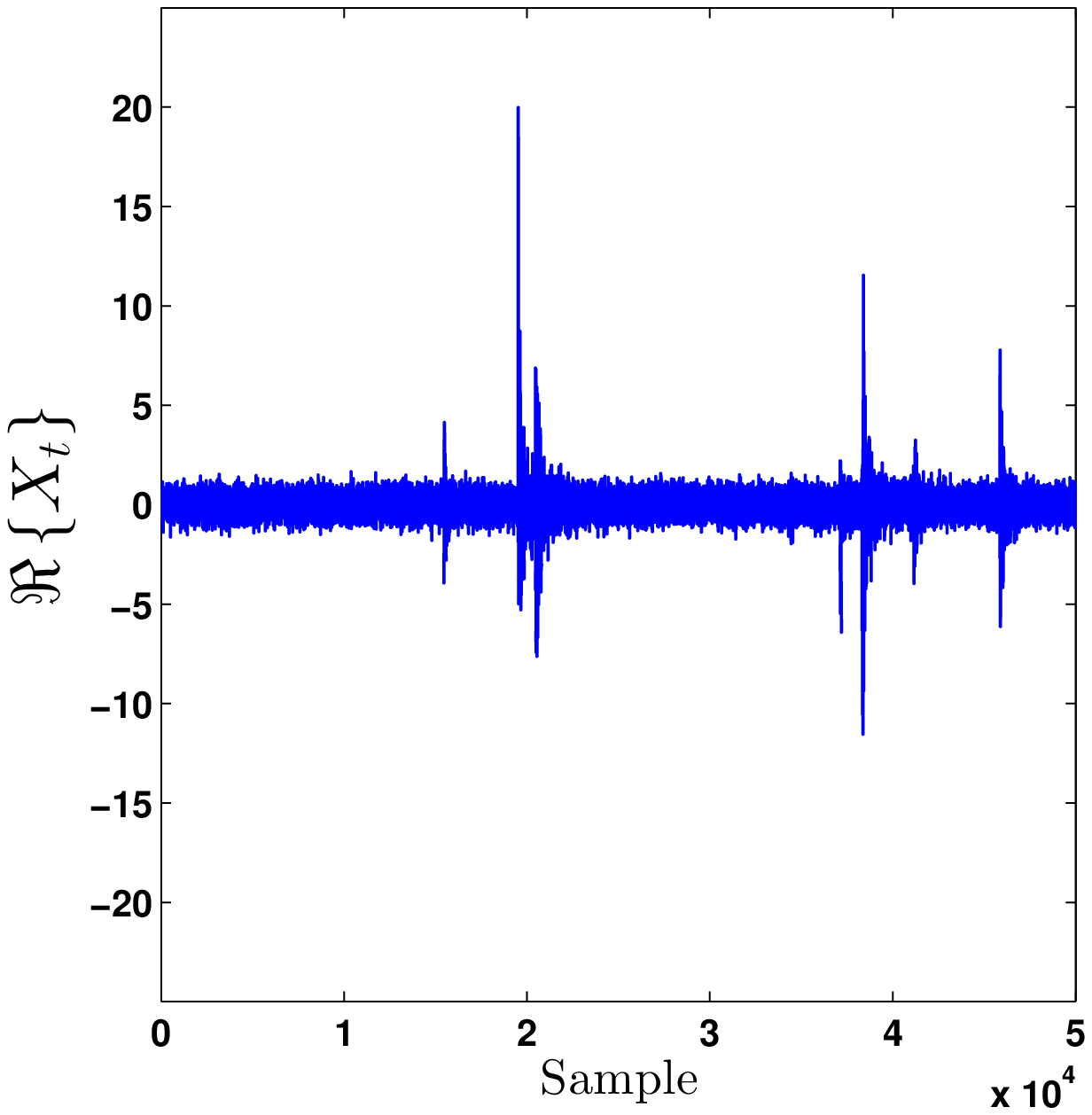}
}
\quad
\subfigure[Amplitude S. $735$ kV  \cite{Au2013} \label{B:VI.B.1}]{
\includegraphics[width=1.48in]{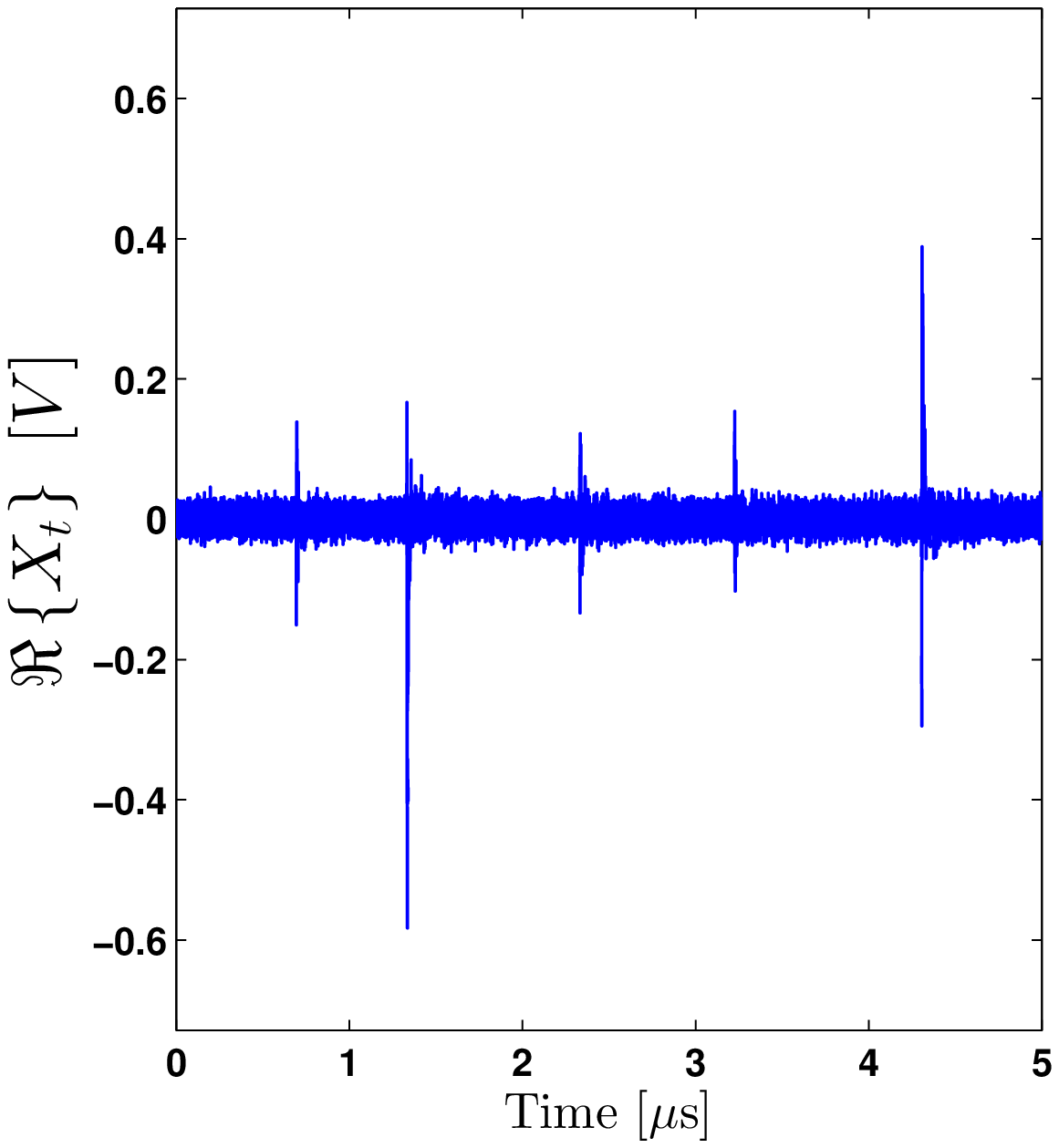}
}
\caption{Example of non-Gaussian process $X_{t}$}
\label{fig:VI.B.1}
\end{figure}

Results of the non-Gaussian noise process from computer simulation and measurement campaign in a $735$ kV substation is provided on Fig. \ref{fig:VI.B.1}. Impulsive noise sample above background noise are produced by impulsive interference sources.  

\subsubsection{Amplitude probability distributions and densities}
\begin{figure}[!htbp]
\centering
\subfigure[Probability density  \label{A:VI.B.2}]{
\includegraphics[width=1.57in]{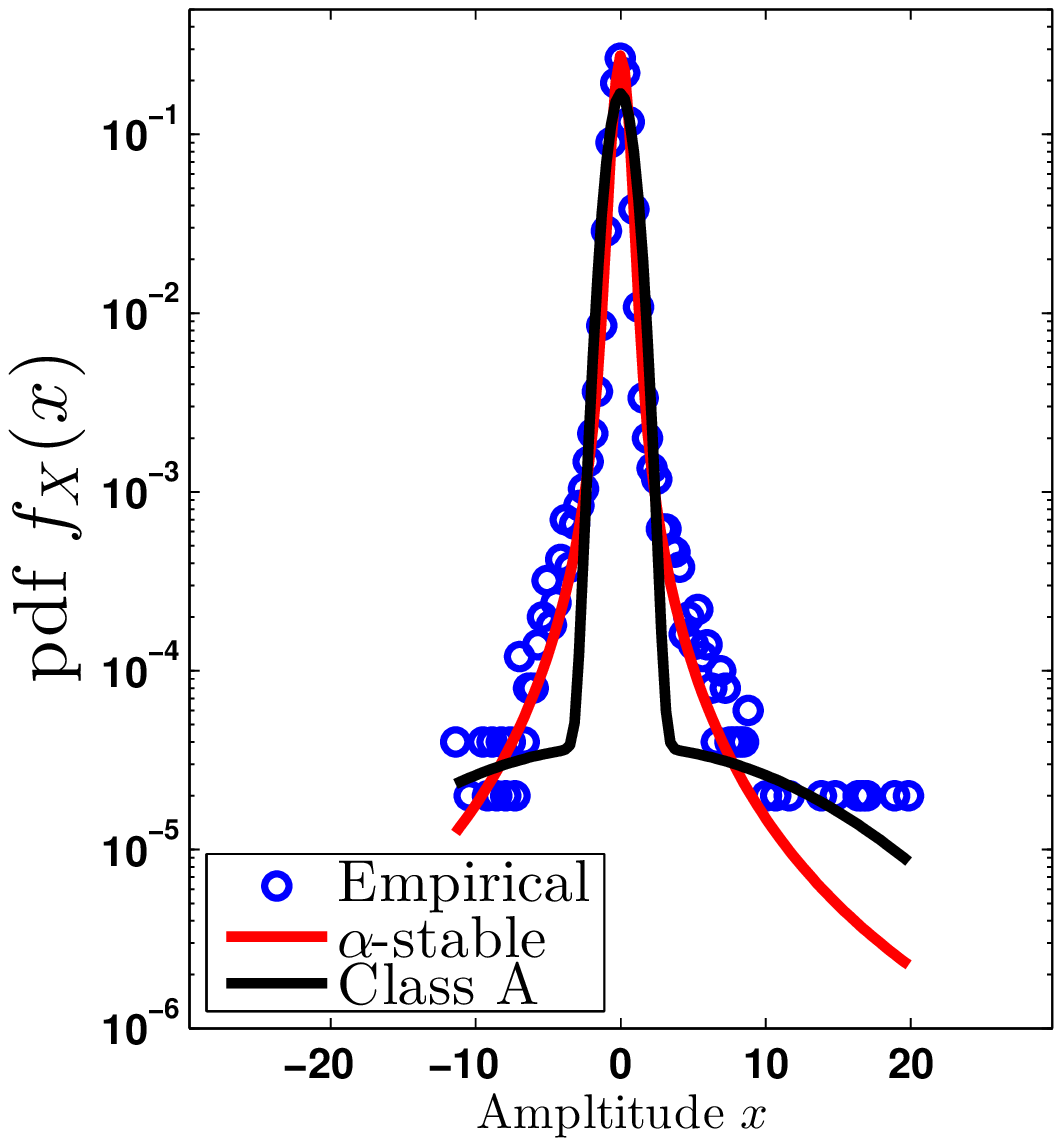}
}
\quad
\subfigure[Tail distribution \label{B:VI.B.2}]{
\includegraphics[width=1.56in]{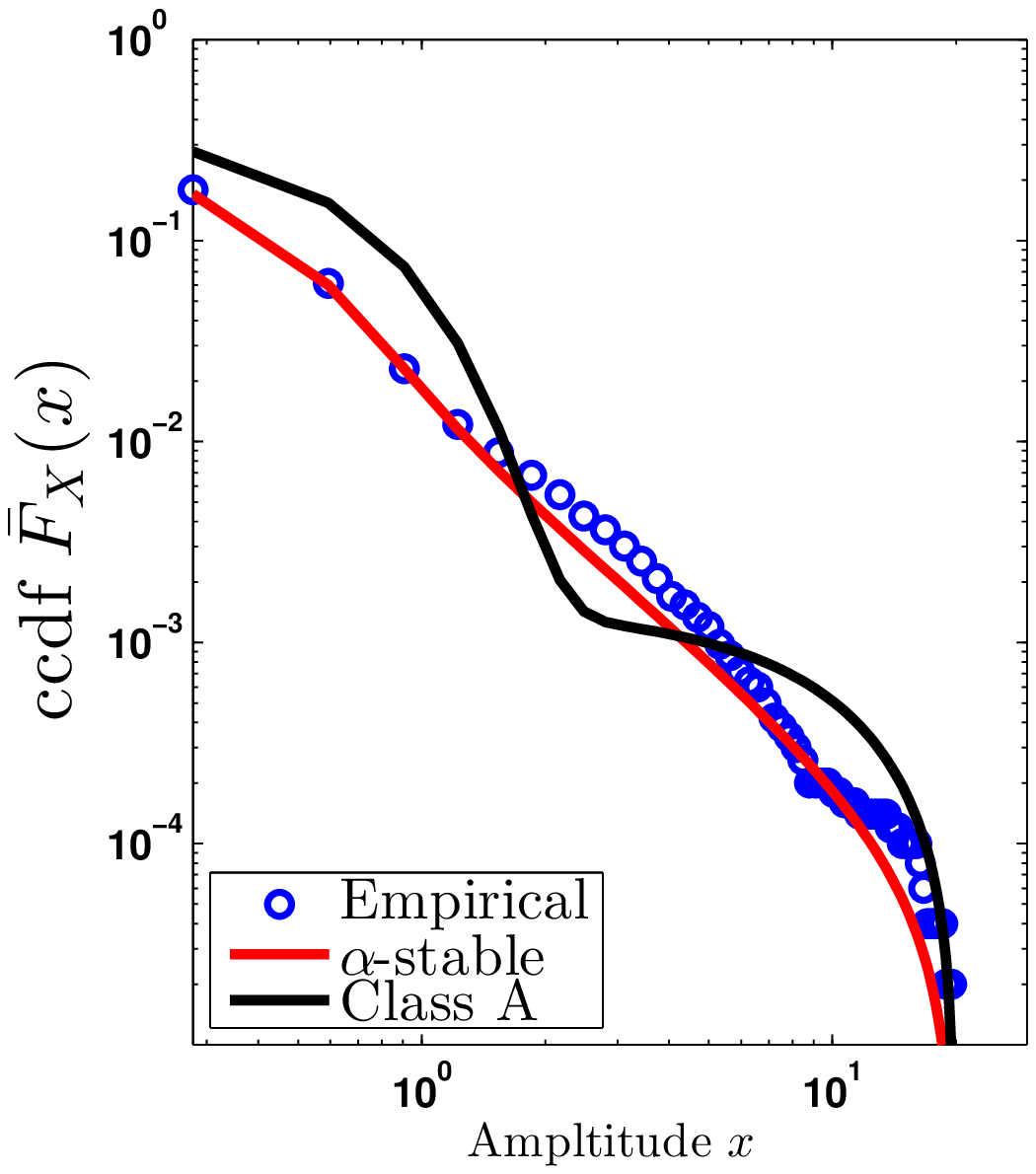}}
\caption{Amplitude distribution and density of non-Gaussian noise process}
\label{fig:VI.B.2}
\end{figure}

\begin{figure}[!htbp]
\centering
\subfigure[Probability density  \label{A:VI.B.3}]{
\includegraphics[width=1.57in]{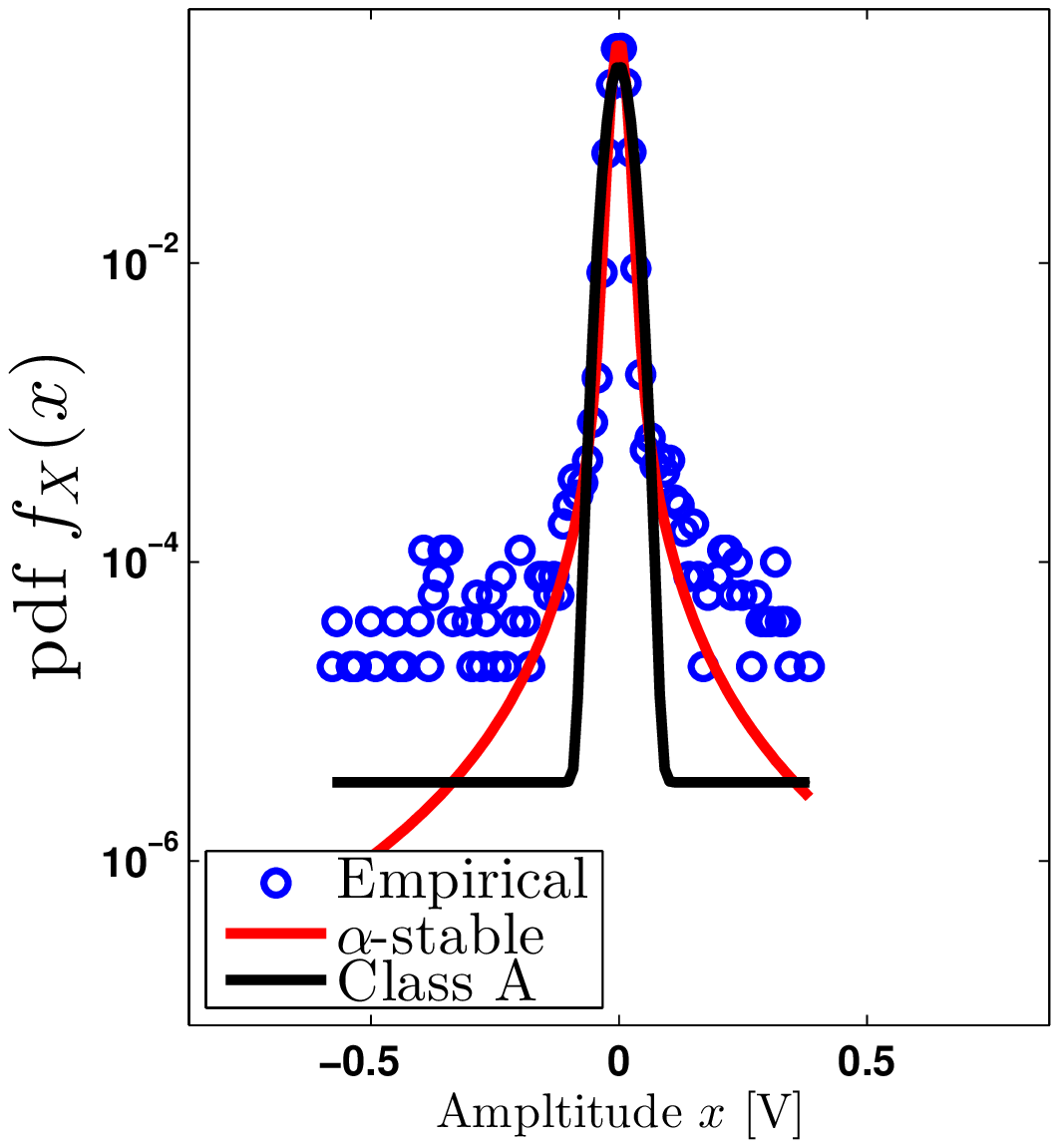}
}
\quad
\subfigure[Tail distribution \label{B:VI.B.3}]{
\includegraphics[width=1.56in]{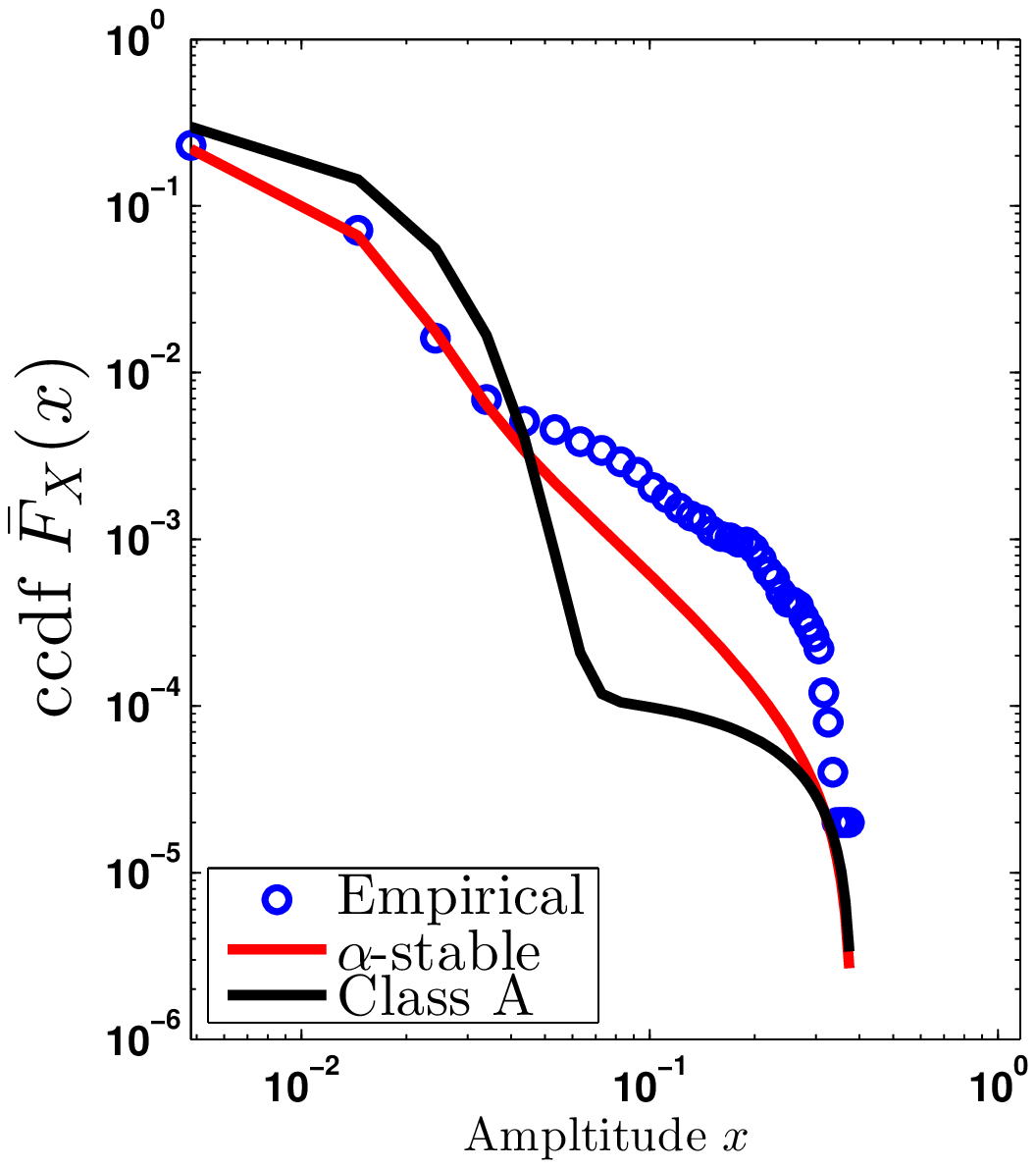}}
\caption{Amplitude distribution and density of non-Gaussian noise measured in $735$ kV substation}
\label{fig:VI.B.3}
\end{figure}

Amplitude distribution and density of non-Gaussian process are depicted on Fig. \ref{fig:VI.B.2} and \ref{fig:VI.B.3} which correspond respectively to samples from the model and from the measurement campaign in a $735$ kV substation. It is seen that the presence of impulsive noise has an influence in terms of amplitude distribution and density. Indeed, low probability of high amplitude can be observed on the tail of the distribution such that we have fat-tailed distributions. This is due to high amplitude of impulsive noise especially when those amplitude distributions are asymptotically power law distributions.   

Classical non-Gaussian noise distributions such as Middleton Class A and $\alpha$-stable distributions have these behaviours, \textit{i.e}, \textit{leptokurtic} distributions and may exhibit an asymmetry as seen on Fig. \ref{fig:VI.B.3}. These two distributions are compared based on empirical data provided by the model and measurements. Parameters of these distributions has been estimated from empirical data based on \cite{Zabin1991,Middleton1979,Koutrouvelis1980,Koutrouvelis1981}. In this condition, amplitude distribution and density has been superposed in order to discuss about the quality of the fit.

From the proposed model and measurements, it can be seen that those distributions can fit empirical data with more or less accuracy. The quality of the fit is determined by the Kullback-Liebler. It is used to measured the divergence of the amplitude density (pdf). Mean square error is used to compare tail distributions (ccdf). Results are set on the table \ref{table:VI.B.2}. It is seen that the $\alpha$-stable distribution fits well the empirical data better than the Middleton Class A. See the KL divergence value and the MSE value of $\alpha$-stable compared to Middleton Class A whatever samples from the model and measurements on the table \ref{table:VI.B.2}. This may be explained by the approximation based on the Edgeworth series expansion where only the first term on the equation (\ref{eq:V.B.14.bis.3}) is used. Therefore, a lack of accuracy is observed in the far tail of distribution. The $\alpha$-stable distribution converge to the empirical data due to the definition given in the equation (\ref{eq:V.B.14.bis.4}) and by arguing that impulsive noise waveforms $U_{t}$ in absolute value can be seen as power law distributions $f_{\vert U\vert}(u) \sim \vert u \vert^{-\alpha-1}$. 

\begin{table}[!htbp]
\caption{Comparison of amplitude distribution and density}
\centering
\renewcommand{\arraystretch}{1.5}
\begin{tabular}{|c|c|c||c|c|}
\hline
& \multicolumn{2}{ |c|| }{Simulation}&\multicolumn{2}{ |c| }{S. $735$ kV}\\
\hline
Model & $\alpha$-stable & Class A & $\alpha$-stable & Class A\\ 
\hline
KL & $0.0037$ & $0.1957$ & $ 0.0111$ & $0.15$\\
\hline 
MSE & $2\cdot 10^{-6} $ & $4.47\cdot 10^{-4} $& $4.16\cdot 10^{-6} $ & $2.37\cdot 10^{-4}$\\
\hline 
\end{tabular} 
\label{table:VI.B.2}
\end{table}

\subsubsection{Power spectral densities} 
\begin{figure}[!htbp]
\centering
\subfigure[Computer simulation \label{A:VI.B.1.1}]{
\includegraphics[width=1.57in]{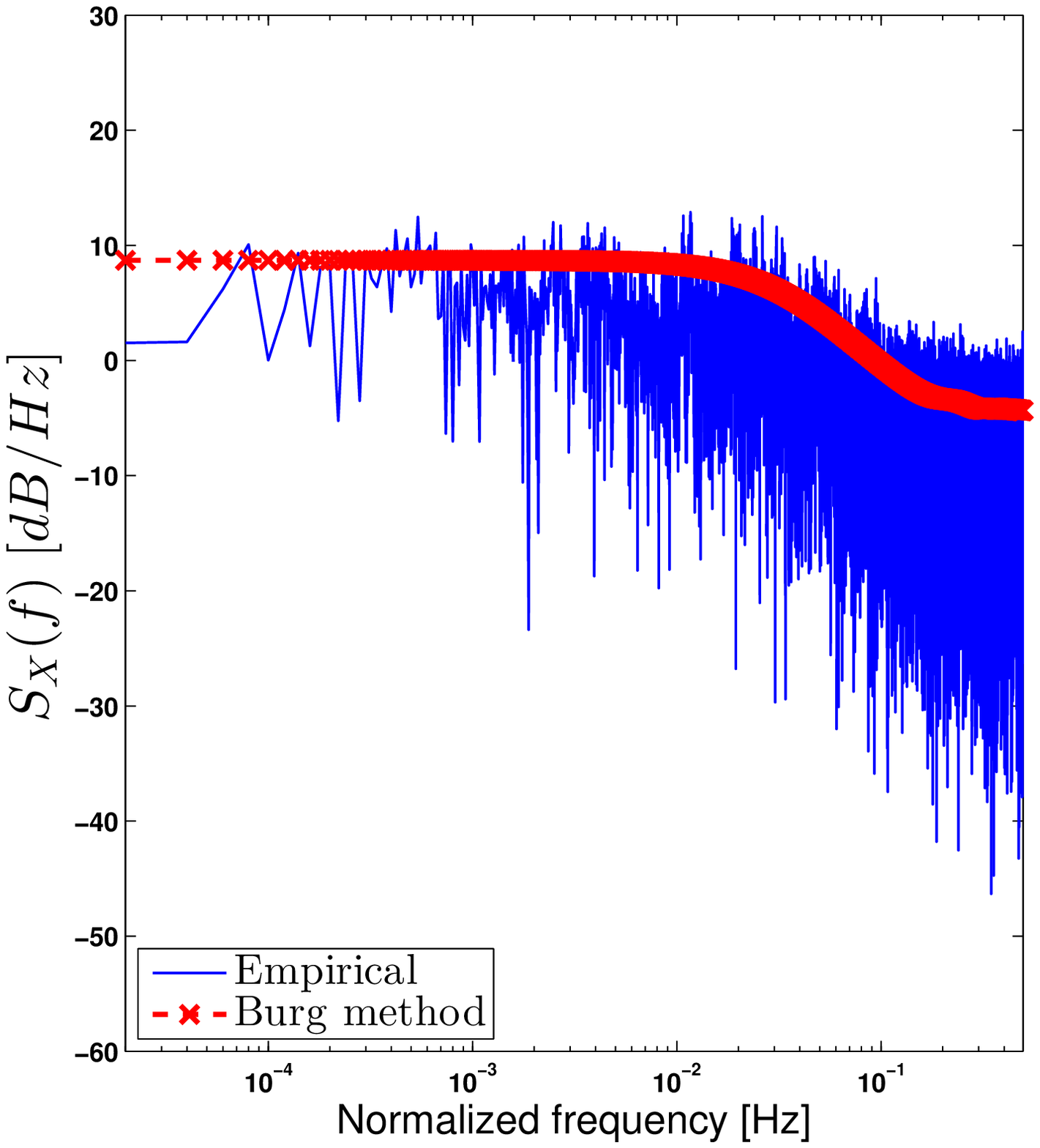}
}
\quad
\subfigure[S. $735$ kV \label{B:VI.B.1.1}]{
\includegraphics[width=1.52in]{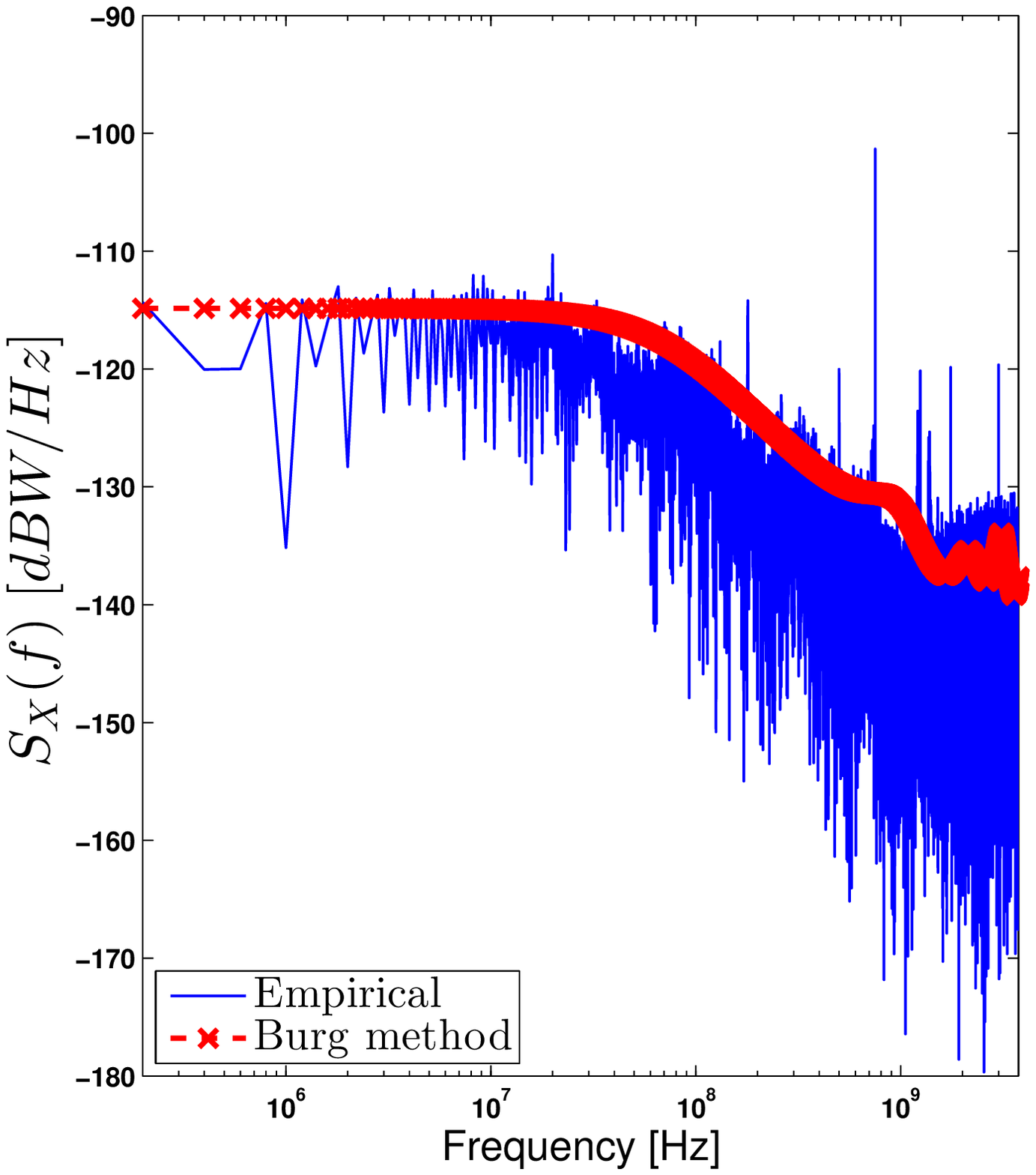}}
\caption{Power spectral density of $X_{t}$}
\label{fig:VI.B.1.1}
\end{figure}

Second order statistics is presented on Fig. \ref{fig:VI.B.1.1}. The psd of the non-Gaussian noise is estimated and smoothed by using parametric method such as Bug's method \cite{Proakis2007,Marple1987} to observe the decay of  $\sim 1/f^{k}$ induced by transient impulsive noise. It is represented by the red curve. On the psd obtained in the $735$ kV substation, wireless communications and harmonics can be observed at $1.5$, $2.5$, $6$ GHz for example. Harmonics are caused by interleaving artefacts and clock feedthrough from scope.

\section{Conclusion}
In this paper we develop a non-Gaussian noise model in presence of transient impulsive noise in substation environments. We use Poisson field of interferers in which impulsive transient interference sources are space-time Poisson process. Based on stochastic geometry, first and second order statistics can be derived. In presence of impulsive noise, it is proved that the amplitude distribution and density can be approximated by classical non-Gaussian noise such as Middleton Class A and $\alpha$-stable distributions. It is seen that the latter is a better approximation than the Middleton Class A due to its approximation by using Edgeworth series expansion. Basic impulsive waveform is specified by using discrete-time series where the innovation process is heteroscedastic to ensure both the the randomness and the transient behaviour of the impulsive interference sources. It is proved that the non-Gaussian noise can be expressed as a second order statistics where the power spectral density is a decay of $\sim 1/f^{k}$. Computer simulation and experimental data are provided to show the validity of the analysis. 

In Future works we will discuss about the reproducibility of the model vis-a-vis the experimentations. Physical parameters such as number of activated interfering sources in the environment, the variance ratio between background noise and the shot-noise process and duration of impulsive noise have to be estimated from the desired environment and validated in terms of first and second order statistics. 

\appendices
\section{Power spectrum density of heteroscedastic process $\varepsilon_{t}$}
\label{Appendix:A}
From the innovation process $\varepsilon_{t}$ defined in equation (\ref{eq:IV.B.1}), we start by calculating the autocorrelation function as :

\begin{equation}
\begin{split}
E\left[ \varepsilon_{t}\varepsilon_{t-k} \right] & = E\left[ \vartheta_{t}\vartheta_{t-k}W_{t}W_{t-k} \right]
\end{split}
 \label{eq:App.A.1}
\end{equation}
The white noise process $W_{t}$ is \textit{i.i.d} such that :

\begin{equation}
\begin{split}
E\left[W_{t}W_{t-k} \right] & = 0
\end{split}
 \label{eq:App.A.2}
\end{equation}
for all values of $k \neq 0$. By assuming that $\vartheta_{t}$ and $W_{t}$ are independent variables, we write the equation (\ref{eq:App.A.1}) as :

\begin{equation}
\begin{split}
E\left[ \varepsilon_{t}\varepsilon_{t-k} \right] & = E\left[ \vartheta_{t}\vartheta_{t}\right]  E\left[W_{t}W_{t} \right]\\
& = E\left[ \vartheta_{t}\vartheta_{t}\right] \sigma_{W}^{2}\delta(0)
\end{split}
 \label{eq:App.A.3}
\end{equation}
where $\sigma_{W}^{2}$ is the variance of the white noise. The Fourier transform of the autocorrelation function of $\varepsilon_{t}$ allow us to write as discrete convolution product between the variance of the white noise and the psd of $\vartheta_{t}$ denoted by $S_{\vartheta}(f)$ :

\begin{equation}
\begin{split}
S_{\varepsilon}(f) & = S_{\vartheta}(f)\ast\sigma_{W}^{2} \\
& = \int_{\mathcal{R}}\sigma_{W}^{2} S_{\vartheta}(f)df\\
& = \sigma_{W}^{2}\sigma_{\vartheta}^{2}\\
\end{split}
 \label{eq:App.A.4}
\end{equation}
where $\sigma_{\vartheta}^{2}$ is the variance of $\vartheta_{t}$. 

\section{About the convergence of the cumulant $\kappa_{m}$ and the cumulant generating function}
\label{Appendix:B}
\subsection{Convergence of $m^{\mbox{\small{th}}}$ cumulant $\kappa_{m}$}
We postulate that $k \leq m$ where $k$ and $m$ are positive integer and $m >0$. From the equation (\ref{eq:V.B.3}), we denote $\left( z_{k}\right) $ a sequence of non-zero real values sequence as :

\begin{equation}
 z_{k} = \lambda {m \choose k}(-1)^{k} \frac{\left\langle  K^{m}\right\rangle }{a(m-k)+b k}
 \label{eq:App.B.1.bis}
 \end{equation} 
By using the ratio test, the convergence of the series $\kappa_{m}$ is ensured if and only if :

\begin{equation}
 L = \lim_{k \rightarrow \infty} \bigg\vert \frac{z_{k+1}}{z_{k}} \bigg\vert < 1
 \label{eq:App.B.1}
 \end{equation}
For the binomial coefficients we have :

\begin{subequations}
\begin{align}
{m \choose 0} = & 1\\
{m \choose k+1} = &{m \choose k} \frac{m-k}{k+1}
\end{align}
\end{subequations} 
In this condition, based on the equation (\ref{eq:V.B.3}), we write $L$ such that :

\begin{equation}
 L = \lim_{k \rightarrow \infty} \bigg\vert  \frac{(-1)(m-k)}{k+1} \bigg\vert  \bigg\vert  \frac{a m +(b-a)(k+1)}{a m +(b-a)k}\bigg\vert 
 \label{eq:App.B.2}
 \end{equation} 
We define $b$ higher than $a$. Moreover, since $k \leq m$ and positive integers, $m$ is necessarily infinite, we determine that $L = 0 < 1$. It is proved that $z_{k}$ is convergent and thus, the series $\kappa_{m}$ is convergent. 

\subsection{The radius of convergence of the cumulant generating function}
From the power series expansion of the cumulant generating function is given by the equation (\ref{eq:V.B.9}), where $s\in \mathcal{C}$ is complex, the radius of convergence of the power series can be discussed. A power series  converge for some values of the variable $s$ and may diverge for others. Thus, the radius of the convergence can be calculated from Cauchy-Hadamard theorem's : 
\begin{equation}
\begin{split}
r^{-1} = \lim_{m \rightarrow \infty} \sup \vert \kappa_{m}\vert^{\frac{1}{m}}
\end{split}
 \label{eq:App.B.4}
 \end{equation}

The radius of the convergence can be calculated from the ratio test of $\kappa_{m}/m!$ :
\begin{equation}
\begin{split}
r^{-1} & = \lim_{m \rightarrow \infty} \bigg\vert  \frac{\kappa_{m+1}}{(m+1) \kappa_{m}}\bigg\vert\\
& =  \lim_{m \rightarrow \infty}   \frac{\sum\limits_{k = 0}^{m+1}{m+1 \choose k}\left[a(m+1-b)+b k \right]\big\vert\left\langle K^{m+1}\right\rangle \big\vert  }{(m+1) \sum\limits_{k = 0}^{m}{m \choose k}\left[a(m-b)+b k \right]\big\vert\left\langle K^{m}\right\rangle \big\vert  }\\
& \sim \lim_{m \rightarrow \infty}  \frac{1}{m+1} = 0
\end{split}
 \label{eq:App.B.5}
 \end{equation}
where $r$ is the radius of convergence which is infinite, $r \rightarrow \infty$, \textit{i.e} the cumulant generating function converges everywhere in the complex plane. Therefore, it is an entire function. 

\section{Power spectral density of $\gamma_{t}$}
\label{Appendix:C}
The power spectral density of $\gamma_{t}$ is given by the Wiener-Khinchine theorem :

\begin{equation}
\begin{split}
 S_{\gamma}(f) & = \int_{\mathcal{R}} E\left[\gamma_{t}\gamma_{t+\tau} \right] e^{j\omega t} dt\\
 S_{\gamma}(f) & = \left\langle K^{2}\right\rangle  \bigg\vert   \frac{1}{\alpha+j\omega}-\frac{1}{b+j\omega}\bigg\vert^{2}\\
 & =   \left\langle K^{2}\right\rangle  \bigg\vert\frac{b-\alpha}{(\alpha+j\omega)(b+j\omega)}\bigg\vert^{2}\\
 &= \frac{\left\langle K^{2}\right\rangle (b-a)^{2}}{(\alpha^{2}+\omega^{2})(b^{2}+\omega^{2})}\\
 \end{split}
\end{equation}
where $\omega = 2\pi f$. The psd of $\gamma_{t}$ is finite such that $ S_{\gamma}(f)< \infty $ for all values of $f\in \mathcal{R}$ by assuming that $\left\langle K^{2}\right\rangle < \infty$. As a result, the integral of the psd  $S_{\gamma}(f) $ is finite.
\ifCLASSOPTIONcaptionsoff
  \newpage
\fi



%
\bibliographystyle{IEEEtran}
\bibliography{Au_Minh_biblio_bdd}

\end{document}